\documentclass[preprint,10pt]{elsarticle}

\usepackage{graphicx}
\usepackage{color}
\usepackage{transparent}
\usepackage{multicol}
\usepackage{float}
\usepackage[font=small,labelfont=small]{caption}
\usepackage{subfig}
\usepackage[section]{placeins}
\usepackage{dblfloatfix}
\usepackage{amssymb}
\usepackage{amsmath}
\usepackage{textcomp}
\usepackage{epstopdf}
\usepackage{ctable}
\usepackage{blindtext}
\usepackage[nodots]{numcompress}
\usepackage{multirow}
\usepackage[margin=3.0cm]{geometry}
\usepackage{setspace}
\usepackage{amsmath}
\biboptions{comma,square,sort&compress}
\usepackage{natbib}

\usepackage{tabulary}
\usepackage{lscape}
\usepackage{adjustbox}
\usepackage{gensymb}
\journal{Composites Part A}

\begin{document}
	\graphicspath{{./Figures/}}
	
	\begin{frontmatter}
		
		\title{A simulation method for fatigue-driven delamination in layered structures involving non-negligible fracture process zones and arbitrarily shaped crack fronts}

		\author[label1,label2]{L. Carreras\corref{cor1}}\ead{lcb@mp.aau.dk Corresponding author}
		\author[label2]{A. Turon} \ead{albert.turon@udg.edu}
		\author[label1]{B.L.V. Bak} \ead{brianbak@mp.aau.dk}	
		\author[label1]{E. Lindgaard} \ead{elo@mp.aau.dk}							
		\author[label2,label3]{J. Renart} \ead{jordi.renart@udg.edu}			
		\author[label4]{F. Martin de la Escalera} \ead{federico.martindelaescalera@aernnova.com}
		\author[label4]{Y. Essa} \ead{yasser.essa@aernnova.com}

		\address[label1]{Dept. of Materials and Production, Aalborg University, Fibigerstraede 16, DK-9220 Aalborg East, Denmark}	
		\address[label2]{AMADE, Polytechnic School, University of Girona, Universitat de Girona 4, E-17003 Girona, Spain}	
		\address[label3]{Serra Húnter Fellow, Generalitat de Catalunya, Spain}				
		\address[label4]{AERNNOVA Engineering Division SA, Llano Castellano Avenue 13, E-28034 Madrid, Spain}
		

\begin{abstract}
	
Most of the existing methods for fatigue-driven delamination are limited to two-dimensional (2D) applications or their predictive capabilities have not been validated in three-dimensional (3D) problems. This work presents a new cohesive zone-based computational method for simulating fatigue-driven delamination in the analysis of 3D structures without crack migration. The method accurately predicts fatigue propagation of non-nelgigible fracture process zones with arbitrarily shaped delamination fronts. The model does not require any kind of fitting parameter since all the input parameters are obtained experimentally from coupon tests. The evaluation of the energy release rate is done using two new techniques recently developed by the authors (the growth driving direction and the mode-decomposed $J$-integral) leading to an accurate prediction of delamination propagation under mixed-mode and non-self-similar growing conditions. The new method has been implemented as a UEL for Abaqus and validated against an experimental benchmark case with varying crack growth rate and shape and extension of the fracture process zone. 

\end{abstract}


\end{frontmatter}


\section{Introduction} \label{sec:Introduction}

The two alternatives to deal with interlaminar fatigue damage in aircraft design are non-growth criterion and the damage tolerance approach  \cite{SAE_International}. Some of the certification processes of aircraft composite structures rely on extensive, and very costly, fatigue testing aimed at ensuring that no growth occurs during the service life. Testing is performed, from small components to large substructures, to evaluate the number of cycles required to make a crack grow perceptibly as a function of the load. On the other hand, recent methods for assessing the certification of aircraft  composite structures \cite{department2006,authority2009,SAE_International}, allow for a ``slow growth" criterion that ensures that any initial delamination inherently present in the structure will grow up to a critical size that can cause the structure collapse during service. If advanced computational tools capable of predicting fatigue damage growth behavior with accuracy are available, an inspection schedule that enables detection of damage before it becomes critical can be reliably programmed. Consequently, a damage tolerance approach can be adopted, enabling lighter and more efficient aircraft designs. 

One of the most recent reviews of simulation methods for delamination propagation \cite{Bakrev} splits them into two categories: linear elastic fracture mechanics (LEFM) and cohesive zone model (CZM) based methods. Another very recent review \cite{Pascoerev}, adds the categories of stress/strain and extended finite element method (XFEM) based models. However, the stress-based expressions are phenomenological models which can be rewritten in terms of LEFM parameters, and most of the XFEM approaches resort to LEFM to define the load. Therefore, attention will be focused on LEFM and CZM based simulation methods.

In general terms, the LEFM based methods \cite{Pradhan1998,Krueger2011a} directly apply any  Paris' law-based expression \cite{Paris1961} for the crack growth rate, $\textrm{d}a/\textrm{d}N$, evaluated using virtual crack closing technique (VCCT) \cite{Krueger2002,Krueger2004}, or any other technique for determining the energy release rate or stress intensity factor. The strength of such methods is that, since they use a phenomenological expression for the crack growth rate, any experimental evidence of the effect of the load ratio, $R$, and/or the mode mixity, $\Phi$, may easily be included in the simulation (provided that the method used in the simulation for characterizing the load is capable of measuring $R$ and $\Phi$). The weaknesses are the assumption of negligible fracture process zone (FPZ), which is not always appropriate, and the need for a pre-existing flaw. An additional limitation related to the VCCT is the requirement of orthogonality of the mesh with the delamination front \cite{Krueger2015}. In contrast, the CZM approach accounts for a non-negligible FPZ, allows to model crack initiation from a pristine interface surface and avoids the need for re-meshing. 

Current cohesive zone models are formulated within the framework of damage mechanics to ensure irreversible crack propagation \cite{Ortiz1999,Alfano2001,Camanho2003,Goyal2004,Turon2006,Turon2010,lindgaard2017user,Simon_multi}. In interface damage mechanics, a damage parameter, $\mathcal{D}$, acts to prevent the restoration of the previous cohesive state between the interfacial surfaces. Most of the existing fatigue formulations are extensions of the cohesive laws for quasi-static loading accounting for an additional criterion for damage development due to fatigue loading. 

The existing CZMs for fatigue delamination can be divided into two main groups \cite{Bakrev}: the loading-unloading hysteresis models \cite{Yang2001,Nguyen2001,Roe2003,Maiti2005,Abdul2005,Tumino2007,Springer2018}, which simulate the whole load cycle to compute the evolution of the damage variable, and the envelope load models \cite{Robinson2005,Turon2007,Pirondi2010,Harper2010,Kawashita2012,Bak2017,Tao2018}, which only model the maximum cyclic load. One of the main advantages of the loading-unloading hysteresis models is that they are capable to model variable loading spectra. In return, as they simulate cycle by cycle, these models become computationally unfeasible for high-cycle fatigue simulation, being only used in low-cycle fatigue applications. Conversely, in envelope load methods, the number of cycles is discretized and the damage variable, $\mathcal{D}$, is updated for each increment in cycles, $\Delta{N}$. The damage at a given number of cycles, $\mathcal{D}(N_{n}+\Delta N )$ is determined by integration of the damage rate, $\textrm{d}\mathcal{D}/\textrm{d}N$. Thus, the envelope load methods are more efficient for simulating high-cycle fatigue.

In \cite{Turon2007}, the authors incorporated a link between the damage rate, $\textrm{d}\mathcal{D}/\textrm{d}N$, and the crack growth rate, $\textrm{d}a/\textrm{d}N$, which was later followed in \cite{Pirondi2010,Harper2010,Kawashita2012,Bak2016,Tao2016,Tao2018}. By using this approach, any influence of the load ratio, $R$, and/or the mode mixity, $\Phi$, is included in the formulation by means of a phenomenological model for the crack growth rate, $\textrm{d}a/\textrm{d}N=f(\mathcal{G}_{max}, R, \Phi...)$. A challenge related to these CZMs is the accurate calculation of the envelope energy release rate, $\mathcal{G}_{max}$, which is needed to evaluate the Paris' law-based expression for $\textrm{d}a/\textrm{d}N$.

The envelope energy release rate, $\mathcal{G}_{max}$, can be computed using the $J$-integral evaluated at the cohesive zone interface \cite{Rice1968}; i.e. integrating the quantities from the cohesive zone model all across the length of the cohesive zone. This approach, used in the methods presented in \cite{Pirondi2010, Moroni2011, Bak2016}, leads to an accurate evaluation of the energy release rate, $\mathcal{G}$. However, since no formulations for the computation of the $J$-integral in 3D modeling of delamination growth, using a cohesive zone model approach, were available at the time they were formulated, these methods were only applicable to 2D simulations. To overcome this limitation, the calculation of the envelope energy release rate, $\mathcal{G}_{max}$, was done by integration of the traction-separation historical response of the most opened point in the cohesive zone, in the methods proposed in \cite{Kawashita2012,Tao2016,Amiri2017}. This implicitly assumes that the historical energy dissipation of the point at the crack tip is comparable to the macroscopic energy release rate, thus, assuming self-similar crack growth. Another approach, used in \cite{Turon2007} and \cite{Harper2010}, is replacing the energy release rate, $\mathcal{G}$, by the total specific work, $\omega_{tot}$, in the Paris' law-based expression for $\textrm{d}a/\textrm{d}N$. The total specific work, $\omega_{tot}$, is the area under the quasi-static law using instantaneous local information at each material point (c.f. Figure \ref{fig:bilinearlaw}). This energy measure, which is a field quantity of the cohesive zone, circumvents the need for the more computationally expensive $J$-integral evaluation of the envelope energy release rate, $\mathcal{G}_{max}$ \cite{Turon_chapter}. 
 
In addition, some of the previous methods make use of the length of the cohesive zone, $l_{cz}$,  in the link between the crack growth rate, $\textrm{d}a/\textrm{d}N$, and the damage rate, $\textrm{d}\mathcal{D}/\textrm{d}N$. The length of the cohesive zone, $l_{cz}$, is a parameter that can be either approximated analytically \cite{Turon2007,Pirondi2010, Moroni2011} or related to that extracted from quasi-static simulations \cite{Harper2010}. Note that the use of fitting parameters that are not possible to be determined experimentally or need to be adjusted depending on the problem make the predictive character and accuracy of the methods questionable. Others \cite{Kawashita2012, Tao2018}, make use of the length associated to an integration point in the crack propagation direction, $l_{e}$, instead. In this case, the propagation direction must be determined in advance. On the other hand, the methods in \cite{Bak2016, Amiri2017} rely on the computation of the spatial derivatives of certain quantities in the CZM formulation along the propagation direction. However, the lack of efficient formulations for the identification of the crack propagation direction prevented these models of being applicable to 3D analysis of delamination.

In \cite{Bak2017}, a benchmark study on six of the aforementioned computational methods \cite{Bak2016,Turon2007,Robinson2005,Pirondi2011,Harper2010,Kawashita2012} is performed. The accuracy of the predicted crack growth rate is studied and compared. Also, the sensitivity to different quasi-static material parameters and method-related fitting parameters are analyzed. It is shown that the method in \cite{Bak2016} is superior in terms of robustness and accuracy. Moreover, it does not include any fitting parameter. It is worth mentioning that, although some of the most recent methods \cite{Tao2016,Amiri2017,Tao2018} were not benchmarked in \cite{Bak2017}, either they are based on any of the previous models or they are not formulated for their applicability to 3D structures.  

In this work, a model capable of predicting delamination growth under both quasi-static and fatigue loading in 3D analysis is presented. It is based on the link between the crack growth rate, $\textrm{d}a/\textrm{d}N$, and the damage rate, $\textrm{d}\mathcal{D}/\textrm{d}N$ presented in \cite{Bak2016}. An efficient methodology to determine the growth driving direction in 3D analysis \cite{Carreras2018} is used. In addition, the model uses a 3D CZ $J$-integral formulation \cite{Carreras_Jint} for calculating the mode-decomposed energy release rates. This 3D CZ $J$-integral formulation takes into account the current loading state over the entire cohesive zone. The model verification is done by simulating the same tests that provided the phenomenological expression for the crack growth rate, $\textrm{d}a/\textrm{d}N$, as input under different mode mixities, $\Phi$, and load ratios, $R$. Finally, the method is validated by comparison of the results obtained from an experimental benchmark test on a partially reinforced double cantilever beam (DCB) specimen with varying crack growth rate and front shape \cite{demo_exp,Carreras_data}. 

\section{Cohesive zone model for quasi-static and fatigue-driven delamination in 3D structures}
\label{sec:CZM}

The point of departure for the formulation developed in this work is the simulation method for high-cycle fatigue-driven delamination presented in \cite{Bak2016}, which was formulated for its applicability to 2D structures. The method in \cite{Bak2016} is, in turn, an extension of the CZM for quasi-static loading proposed in \cite{Turon2006,Turon2010}. 

The quasi-static cohesive model is formulated in the framework of damage mechanics. A thermodynamical representation of the irreversible process of interlaminar fracture is done by defining an energy-based damage variable, $\mathcal{D}^{e}$. It corresponds to the fraction of dissipated specific energy, $\omega_{d}$, to the energy necessary to create a unit of new surface under quasi-static loading conditions, which is the fracture toughness, $\mathcal{G}_{c}$ (c.f. Figure \ref{fig:bilinearlaw}). The fatigue model is based on an envelope load approach, in which the energy-based damage rate, $\textrm{d}\mathcal{D}^{e}/\textrm{d}N$, is directly related to the crack growth rate, $\textrm{d}a/\textrm{d}N$, avoiding making use of any fitting parameter. This is achieved by maintaining the quasi-static relationship between tractions, $\boldsymbol{\tau}$, and displacement jumps, $\boldsymbol{\delta}$, during fatigue propagation \cite{Bak2016}.

\subsection{Quasi-static damage model}
\label{sec:CZM_sta}

\begin{figure}[h]
	\centering
	\includegraphics[width=8 cm]{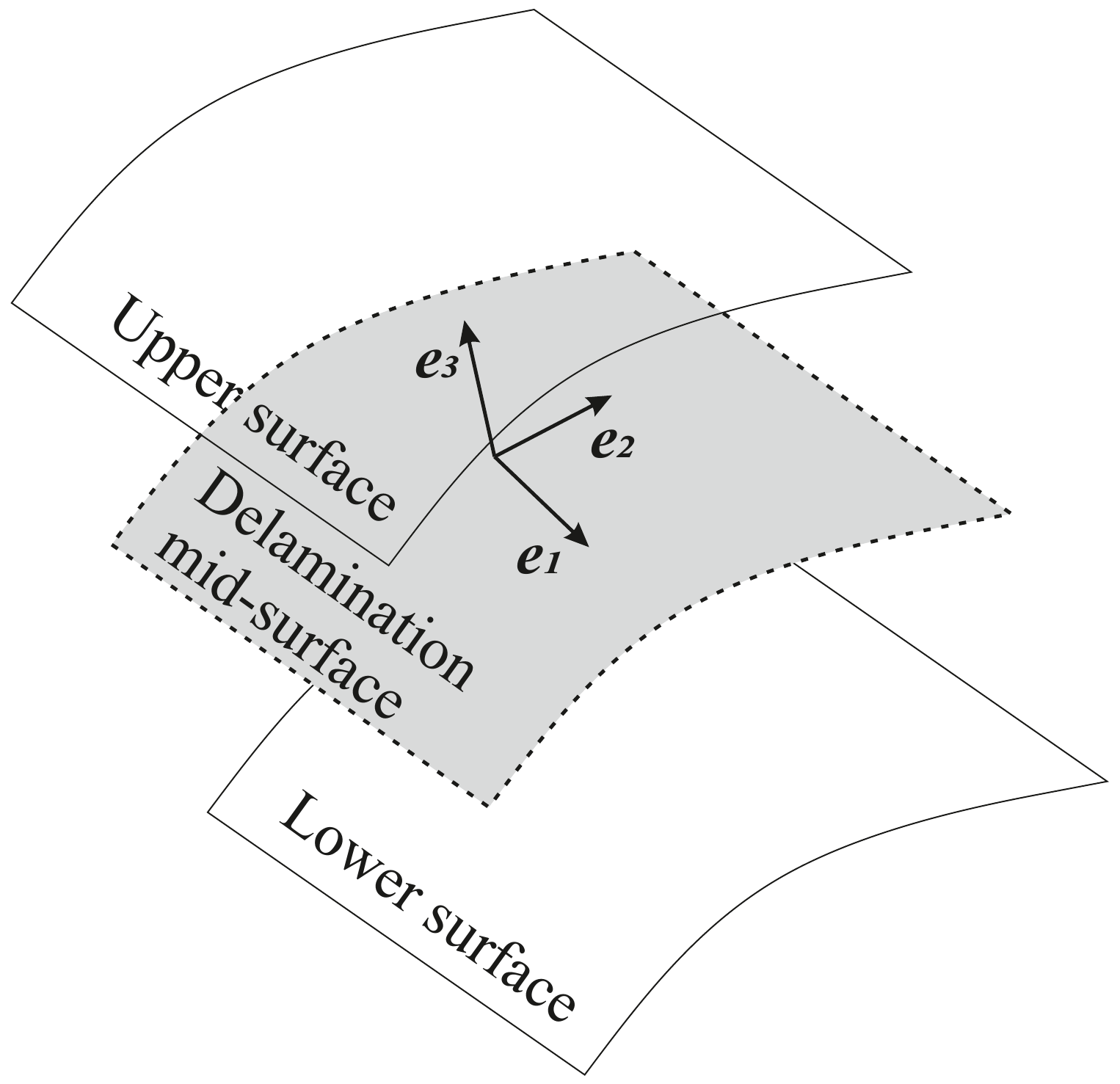} 	
	\caption{Local Cartesian coordinate system $(e_{1},e_{2},e_{3})$ on the delamination mid-surface.}
	\label{fig:midsurface}	
\end{figure}

The CZM in \cite{Turon2006,Turon2010} describes the cohesive behavior of a band of micro-cracked material ahead of the crack tip capable of transferring stress until complete separation. The distribution of cohesive tractions, $\boldsymbol{\tau}(\boldsymbol{\delta})$, is a function of the separations between crack faces, $\boldsymbol{\delta}$, referred to as displacement jumps. Defining a local Cartesian coordinate system on the delamination mid-surface such that $\boldsymbol{e_{1}}$ and $\boldsymbol{e_{2}}$ are the local tangential directions  and  $\boldsymbol{e_{3}}$ is the local normal direction (c.f. Figure \ref{fig:midsurface}), the constitutive relation is defined as:  

\begin{equation}
\begin{split}
\tau_{j}&=\left(1-\mathcal{D}^{K}\right) K \delta_{j} \qquad \text{for} \qquad j=1,2 \\  
\tau_{3}&=\left(1-\mathcal{D}^{K}\right) K \delta_{3}-\mathcal{D}^{K} K \langle-\delta_{3}\rangle
\label{eq:constitutiverelaction}
\end{split}
\end{equation}

\noindent where $\mathcal{D}^{K} \in [0,1]$ is a scalar damage variable reducing the penalty stiffness, $K$, introduced in the numerical implementation of the CZM in a finite element framework.

\begin{figure}[h]
	\centering
	\includegraphics[width=8 cm]{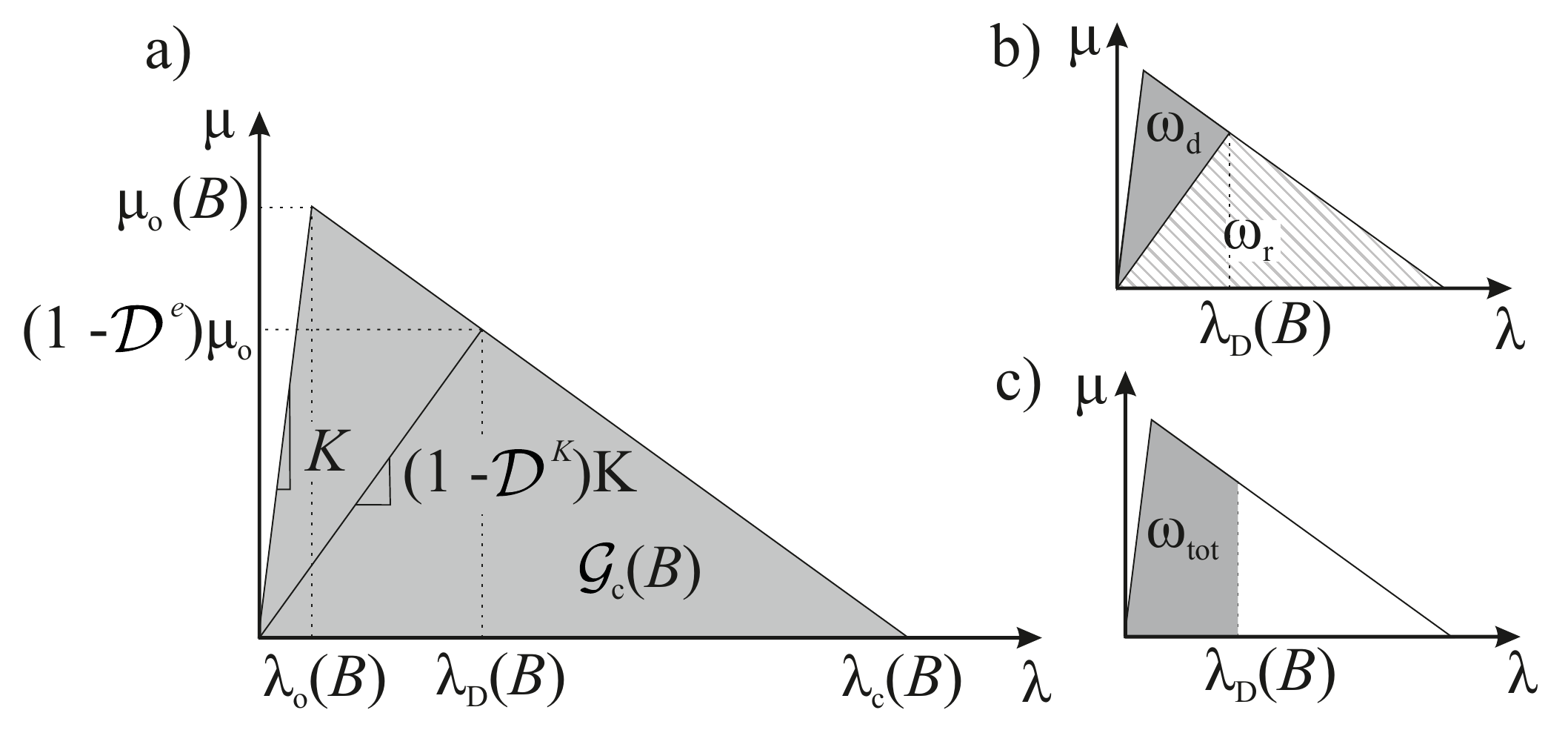} 	
	\caption{Equivalent one-dimensional cohesive law for a given mode-mixity, $B$. The shadowed area in a) represents the fracture toughness, $\mathcal{G}_{c}$, in b), the specific dissipated energy, $\omega_{d}$, and the specific remaining ability to do non-conservative work, $\omega_{r}$, and in c), the total specific work, $\omega_{tot}$, for a given state of damage.}
	\label{fig:bilinearlaw}	
\end{figure}

The evolution of the stiffness degrading damage variable, $\mathcal{D}^{K}$, is governed by an equivalent one-dimensional cohesive law (c.f. Figure \ref{fig:bilinearlaw}) to ensure irreversibility under changing mixed-mode loading conditions. The equivalent one-dimensional displacement jump is defined as:

\begin{equation}
\lambda=\sqrt{\left(\delta_{I} \right)^2+\left( \delta_{s}\right)^2}
\label{eq:lambda} 
\end{equation}

\noindent where $\delta_{I}$ is the displacement jump associated to mode I and  $\delta_{s}$ is the shear sliding resultant of the displacement jumps associated to any combination of mode II and mode III.

\begin{equation}
\delta_{I}=\langle \delta_{3}\rangle, \qquad  \delta_{s}=\sqrt{\left(\delta_{1} \right)^2+\left( \delta_{2}\right)^2}
\label{eq:displjump} 
\end{equation}   

The process of fracture initiates when the stress reaches the interlaminar strength, $\mu_{o}$. Then, the cohesive law describes a stress-softening behavior. With increasing displacement jump, $\lambda$, cohesive traction, $\mu$, decreases. The amount of energy dissipated per unit of newly created crack area in the opening process (from 0 to the critical displacement jump, $\lambda_{c}$) is the fracture toughness, $\mathcal{G}_{c}$. The fracture toughness, $\mathcal{G}_{c}$, for a given mode-mixity, is determined by \cite{Benzeggagh1996}:

\begin{equation}
\mathcal{G}_{c}=\mathcal{G}_{Ic}+\left(\mathcal{G}_{sc}-\mathcal{G}_{Ic} \right)B^{\eta}
\label{eq:BK} 
\end{equation}

\noindent where subscripts $Ic$ and $sc$ denote the pure mode I and shear quasi-static critical values, respectively, and $\eta$ is a mode interaction parameter determined by fitting Equation (\ref{eq:BK}) to a batch of experimentally measured fracture toughnesses under different mode-mixities. Note that Equation (\ref{eq:BK}) was originally formulated in \cite{Benzeggagh1996} for the mode I-II interpolation of the fracture toughness and, thus, $\mathcal{G}_{IIc}$ was used instead of $\mathcal{G}_{sc}$. In the applied CZM \cite{Turon2006,Turon2010}, modes II and III are not disregarded and the shear properties are treated as mode II ($\mathcal{G}_{sc}=\mathcal{G}_{IIc}=\mathcal{G}_{IIIc}$). $B$ is the local displacement-based mode-mixity and is defined in terms of the displacement jump as:

\begin{equation}
B=\frac{\delta_{s}^{2}}{\delta_{I}^{2}+\delta_{s}^{2}}
\label{eq:B} 
\end{equation}

The mixed-mode intepolation of the interlaminar strength, $\mu_{o}$, is done by:

\begin{equation}
\mu_{o}=\sqrt{\left(\tau_{Io}\right) ^2+\left[\left(\tau_{so})^2-(\tau_{Io} \right)^2\right] B^{\eta}}
\label{eq:BK_mu} 
\end{equation}

\noindent where $\tau_{Io}$ and $\tau_{so}$ are the tensile and shear interfacial strengths, where $\tau_{so}=\tau_{IIo}=\tau_{IIIo}$. A mode-independent penalty stiffness, $K$, is used which sets the interfacial shear strength dependent on the other material properties \cite{Turon2010}:

\begin{equation}
\tau_{so}=\tau_{Io}\sqrt{\frac{\mathcal{G}_{sc}}{\mathcal{G}_{Ic}}}
\label{eq:tauIo_tauIIo} 
\end{equation}

The onset, $\lambda_{o}$, and propagation, $\lambda_{c}$, of delamination in terms of  the equivalent one-dimensional displacement jump are expressed as:

\begin{equation}
\lambda_{o}=\frac{\mu_{o}}{K}, \qquad  \lambda_{c}=\frac{2 \mathcal{G}_{c}}{\mu_{o}}
\label{eq:lambda_o_and_c} 
\end{equation}

Finally, the stiffness degrading, $\mathcal{D}^{K}$, and the energy-based, $\mathcal{D}^{e}$, damage variables are related by:

\begin{equation}
\mathcal{D}^{e}=1-\frac{\left( 1-\mathcal{D}^{K}\right)  \lambda_{\mathcal{D}} }{\lambda_{o}}
\label{eq:DeDk} 
\end{equation}

\noindent where $\lambda_{\mathcal{D}}$ is the equivalent one-dimensional displacement jump associated to the current damage state, defined as:

\begin{equation}
\lambda_{\mathcal{D}}=\frac{\lambda_{o} \lambda_{c}}{\lambda_{c}-\mathcal{D}^{K}\left(\lambda_{c}-\lambda_{o} \right) }
\label{eq:lambda_D} 
\end{equation}

\subsection{Fatigue damage rate model}
\label{sec:CZM_fat}

Since $\mathcal{D}^{e}$ depends on the local displacement-based mode-mixity, $B$, and the equivalent one-dimensional displacement jump, $\lambda$, the rate of energy-based damage, $\frac{\textrm{d}\mathcal{D}^{e}}{\textrm{d} N}$, is deduced by applying the chain rule \cite{Bak2016} as:

\begin{equation}
\frac{\textrm{d}\mathcal{D}^e}{\textrm{d} N}=\left(\frac{\partial\mathcal{D}^e}{\partial B} \frac{\partial B}{\partial a}+\frac{\partial \mathcal{D}^e}{\partial \lambda} \frac{\partial\lambda}{\partial a} \right)\frac{\textrm{d}a}{\textrm{d} N}  
\label{eq:dDedN}
\end{equation}

Each of the partial derivatives in Equation (\ref{eq:dDedN}) are addressed in the following. The derivative of the energy-based damage with respect to the mode mixity, $\frac{\partial \mathcal{D}^{e}}{\partial B}$, and the derivative of the energy-based damage with respect to the equivalent one-dimensional displacement jump, $\frac{\partial \mathcal{D}^{e}}{\partial \lambda}$, are derived once again applying the chain rule:

\begin{equation}
\begin{split}
\frac{\partial\mathcal{D}^{e}}{\partial B}&= \frac{\partial\left(\frac{\omega_{d}}{\mathcal{G}_{c}} \right)  }{\partial \omega_{d}} \left( \frac{\partial \omega_{d}}{\partial \lambda_{o}}\frac{\partial \lambda_{o}}{\partial \mu_{o}} +  \frac{\partial \omega_{d}}{\partial \lambda_{c}}\frac{\partial \lambda_{c}}{\partial \mu_{o}}  \right) \frac{\partial \mu_{o}}{\partial B} +\left( \frac{\partial\left(\frac{\omega_{d}}{\mathcal{G}_{c}} \right)  }{\partial \omega_{d}}  \frac{\partial \omega_{d}}{\partial \lambda_{c}}   \frac{\partial \lambda_{c}}{\partial \mathcal{G}_{c}}  +   \frac{\partial\left(\frac{\omega_{d}}{\mathcal{G}_{c}} \right)  }{\partial \mathcal{G}_{c}} \right) \frac{\partial \mathcal{G}_{c}}{\partial B} \\
\frac{\partial\mathcal{D}^{e}}{\partial \lambda}&= \frac{\partial\left(\frac{\omega_{d}}{\mathcal{G}_{c}} \right)  }{\partial \omega_{d}} \frac {\partial \omega_{d}}{\partial \lambda}
\end{split}
\label{eq:dDedBdlambda}
\end{equation}

The partial derivatives in Equation (\ref{eq:dDedBdlambda}) are solved for the particular application of the CZM from \cite{Turon2006, Turon2010} and listed in Table \ref{tab:Derivatives}. The expressions for $\frac{\partial \mathcal{D}^{e}}{\partial B}$ and  $\frac{\partial \mathcal{D}^{e}}{\partial \lambda}$, obtained after substituting the partial derivatives in Equation (\ref{eq:dDedBdlambda}) by the expressions listed in Table \ref{tab:Derivatives} read:

\begin{equation}
\begin{split}
\frac{\partial \mathcal{D}^{e}}{\partial B}&=\frac{\eta \left(\mathcal{G}_{sc}-\mathcal{G}_{Ic}  \right)B^{\left(\eta-1 \right) } \lambda}{K \lambda_{c} \lambda_{o} \left( \lambda_{o}-\lambda_{c}\right) }\\
\frac{\partial\mathcal{D}^{e}}{\partial\lambda}&=\frac{1}{\lambda_{c}-\lambda_{o}}
\end{split}
\label{eq:FB}
\end{equation}

\begin{table}[]
	\centering

	\begin{tabular}{lll}
		\hline \hline
		Dependencies          & Partial derivatives &  \\ \hline \\
		\multirow{1}{*}{$\displaystyle{\lambda_{o}\left( \mu_{o}\right) }$}   &$\displaystyle{\frac{ \partial \lambda_{o}}{\partial  \mu_{o}}=\frac{1}{K}} $            & \refstepcounter{equation}(\theequation \label{eq:dlambdao_dmuo})   
		\\ \\
		\multirow{2}{*}{$\displaystyle{\lambda_{c}\left( \mu_{o}, \mathcal{G}_{c}\right)} $} & $\displaystyle{\frac{ \partial \lambda_{c}}{\partial  \mu_{o}}=-\frac{2\mathcal{G}_{c}}{\mu_{o}^{2}}} $      &  \refstepcounter{equation}(\theequation \label{eq:dlambdac_dmuo})       \\
		& $\displaystyle{\frac{ \partial \lambda_{c}}{\partial  \mathcal{G}_{c}}=\frac{2}{\mu_{o}}} $      &   \refstepcounter{equation}(\theequation \label{eq:dlambdac_dGc})      \\ \\
		\multirow{1}{*}{$\displaystyle{\mu_{o}\left( B\right)} $}   & $\displaystyle{\frac{ \partial \mu_{o}}{\partial B}=\frac{\eta\left(\tau_{so}^2- \tau_{Io}^2\right) B^{\eta-1}}{2\mu_{o}}} $            & \refstepcounter{equation}(\theequation \label{eq:dmuo_dB})   
		\\ \\
		\multirow{1}{*}{$\displaystyle{\mathcal{G}_{c}\left( B\right)} $}   & $\displaystyle{\frac{ \partial \mathcal{G}_{c}}{\partial B}=\eta\left(\mathcal{G}_{sc}- \mathcal{G}_{Ic}\right) B^{\eta-1} }$            & \refstepcounter{equation}(\theequation \label{eq:dGc_dB})   
		\\ \\	
		\multirow{3}{*}{$\displaystyle{\omega_{d}\left( \lambda_{o}, \lambda_{c}, \lambda \right) }$} & $\displaystyle{\frac{\partial \omega_{d}}{\partial \lambda_{o}}=\frac{1}{2} K \lambda_{c} \frac{ \lambda_{o}^{2}-2\lambda_{c} \lambda_{o}+ \lambda_{c} \lambda }{ \left( \lambda_{o}-\lambda_{c}\right) ^{2}}}$       &  \refstepcounter{equation}(\theequation \label{eq:dwd_dlambdao})       \\
		&$\displaystyle{\frac{\partial \omega_{d}}{\partial \lambda_{c}}=\frac{1}{2} K \lambda_{o}^{2} \frac{ \lambda_{o}- \lambda }{ \left( \lambda_{o}-\lambda_{c}\right) ^{2}}}$    &   \refstepcounter{equation}(\theequation \label{eq:dwd_dlambdac})      \\
		& $\displaystyle{\frac{ \partial \omega_{d}}{\partial  \lambda}=\frac{1}{2}K \lambda_{o}\lambda_{c} \frac{1}{\left(\lambda_{c}-\lambda_{o} \right)}} $    &   \refstepcounter{equation}(\theequation \label{eq:dwd_dlambda})     			
		\\ 	\\
		\multirow{2}{*}{$\displaystyle{\mathcal{D}^{e}\left( \omega_{d}, \mathcal{G}_{c}\right) }$} & $\displaystyle{\frac{ \partial \left( \frac{\omega_{d}}{\mathcal{G}_{c}}\right) }{\partial  \omega_{d}}=\frac{1}{\mathcal{G}_{c}}} $      &  \refstepcounter{equation}(\theequation \label{eq:dwdGc_dwd})       \\
		& $\displaystyle{\frac{ \partial \left( \frac{\omega_{d}}{\mathcal{G}_{c}}\right) }{\partial  \mathcal{G}_{c}}=\frac{-\omega_{d}}{\mathcal{G}_{c}^2}} $      &   \refstepcounter{equation}(\theequation \label{eq:wdGc_dGc})   
		\\		\\	 		
		
		\hline \hline    
	\end{tabular}
		\caption{Dependencies and partial derivatives of the variables in the system using the CZM presented in \cite{Turon2006,Turon2010}.}
	\label{tab:Derivatives}
\end{table}

In a self-similar crack growth, $\frac{\partial B}{\partial a}$ and $\frac{\partial\lambda}{\partial a}$ can be interpreted as the slopes of the local mode-mixity and the equivalent one-dimensional displacement jump along the direction of crack propagation (for further details on this assumption, the reader is referred to the original paper \cite{Bak2016} describing the 2D method). Thus, if $x_{1}$ is defined as the crack growth direction coordinate, the rates $\frac{\partial B}{\partial a}$ and $\frac{\partial\lambda}{\partial a}$ may be approximated by:

\begin{equation}
\begin{split}
\frac{\partial B}{\partial a} &\approx \frac{\partial B}{\partial x_{1}} \\ \frac{\partial \lambda}{\partial a} &\approx\frac{\partial \lambda}{\partial x_{1}}
\end{split}
\label{eq:slopes}
\end{equation}

Then, by application of the chain rule, $\frac{\partial B}{\partial x_{1}}$ and $\frac{\partial \lambda}{\partial x_{1}}$  read:

\begin{equation}
\begin{split}
\frac{\partial B}{\partial x_{1}}=\frac{\partial B}{\partial \delta_{j}} \frac{\partial \delta_{j}}{\partial x_{1}}\\
\frac{\partial \lambda}{\partial x_{1}}=\frac{\partial \lambda}{\partial \delta_{j}} \frac{\partial \delta_{j}}{\partial x_{1}}\\
\end{split}
\label{eq:dB_ddelta_ddelta_dx1}
\end{equation}

\noindent where the terms $\frac{\partial B}{\partial \delta_{j}}$ and  $\frac{\partial \lambda}{\partial \delta_{j}}$ are solved as:

\begin{equation}
\begin{split}
\frac{\partial B}{\partial \delta_{1}}&=\frac{2 \delta_{1} \langle \delta_{3}\rangle^2}{\lambda^4};   \qquad  \frac{\partial B}{\partial \delta_{2}}=\frac{2 \delta_{2} \langle \delta_{3}\rangle^2}{\lambda^4};  \qquad  \frac{\partial B}{\partial \delta_{3}}=-\frac{2 \delta_{s}^2 \langle \delta_{3}\rangle }{\lambda^4}\\
\frac{\partial \lambda}{\partial \delta_{1}}&=\frac{\delta_{1}}{\lambda };  \qquad \qquad \frac{\partial \lambda}{\partial \delta_{2}}=\frac{\delta_{2}}{\lambda};  \qquad \qquad \frac{\partial \lambda}{\partial \delta_{3}}=\frac{\langle \delta_{3}\rangle}{\lambda}
\end{split}
\label{eq:dlambda_ddelta}
\end{equation}

\begin{figure}[h!]
	\centering
	\includegraphics[width=8 cm]{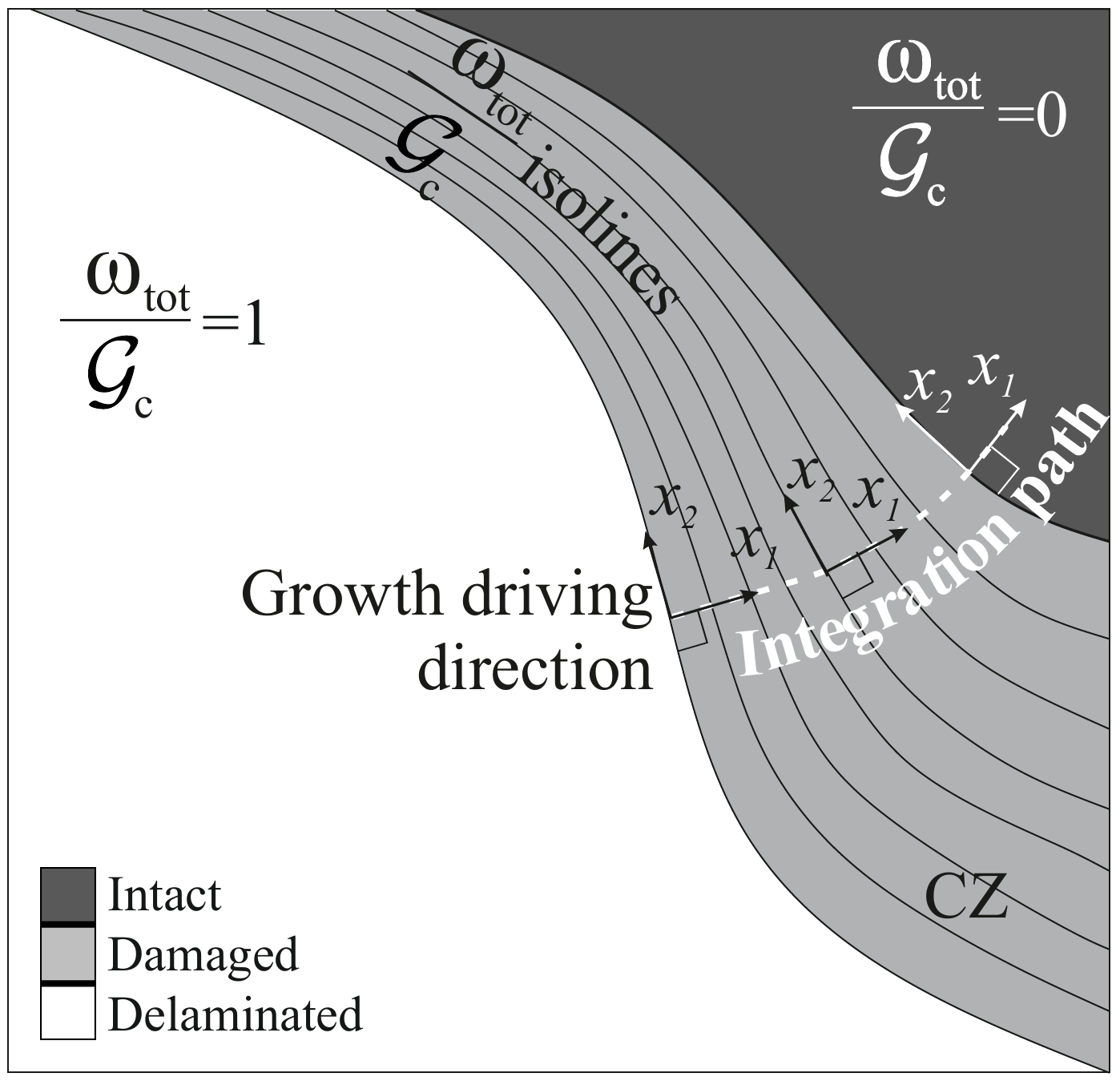} 	
	\caption{Schematic representation of the delamination surface. The growth driving direction (GDD) is assumed to be the normal direction to $\frac{\omega_{tot}}{\mathcal{G}_{c}}$ isolines. The local crack coordinate system is defined by $x_{1}$ tangent to GDD, $x_{2}$ normal to GDD and $x_{3}$ normal to the delamination surface. The integration paths of the 3D CZ $J$-integral are tangent to the local GDD direction. }
	\label{fig:sketch_GDD}	
\end{figure}

Note that, in order to compute $\frac{\partial \delta_{j}}{\partial x_{1}}$ of Equation (\ref{eq:dB_ddelta_ddelta_dx1}), the local Cartesian coordinates $(e_{1}, e_{2}, e_{3})$ must be rotated around $e_{3}$ (c.f. Figure \ref{fig:midsurface}) so that a new local Cartesian coordinate system $(x_{1}, x_{2}, x_{3})$ is defined, where $x_{1}$ is the direction of crack propagation and tangent to the mid-surface, $x_{3}$ is the normal direction to the mid-surface and $x_{2}$ is perpendicular to $x_{1}$ and $x_{3}$. Therefore, the propagation direction must be identified. In the framework of LEFM, the propagation direction is usually understood as the direction normal to the crack front. Conversely, if a CZM approach is used, the crack front is not defined by a line, but there is a band of damaged material ahead of the crack tip, and thus the normal direction to a front line does not apply. In this case, the concept of growth driving direction (GDD) \cite{Carreras2018} can be used. The GDD can be defined as the gradient vector field of the total specific work, $\omega_{tot}$, over the fracture toughness, $\mathcal{G}_{c}$, with respect to the coordinates tangent to the mid-surface (c.f. Figure \ref{fig:sketch_GDD}). It can be efficiently evaluated at any point within the cohesive zone using information available at finite element level. Moreover, the formulation presented in \cite{Carreras2018} also addresses the calculation of the slopes $\frac{\partial \delta_{j}}{\partial x_{1}}$, although the main steps are introduced in the following for completeness. By assuming that the curvature of the interface within the element domain is small, the derivatives $\frac{\partial \delta_{j}}{\partial x_{1}}$ read:

\begin{equation}
\frac{\partial \delta_{j}}{\partial x_{1}}= R_{ji}  \left[ -\frac{\partial N_{ik }}{\partial \eta_{\alpha}}, \  \frac{\partial N_{ik }}{\partial \eta_{\alpha}} \right]  \left[ R_{1p} \frac{1}{2}\frac{\partial N_{pm}}{\partial \eta_{\alpha}}\left(C_{m}^{-} +C_{m}^{+}+ Q_{m}^{-}+ Q_{m}^{+}\right)\right]^{-1}   \left[ Q_{k}^{-}, Q_{k}^{+} \right]
\label{eq:ddelta_de1}
\end{equation}

\noindent where $R_{ji}$ is the transformation tensor that relates the global coordinate $i$ to the local crack coordinate $j$. $\frac{\partial N_{ik}}{\partial \eta_{\alpha}}$ is the derivative of the shape function matrix with respect to the $\alpha$-isoparametric coordinate of the finite element, in which subscript $k$ runs from 1 to the number of degrees of freedom of each of the surfaces of the cohesive element (lower and upper).  $C_{m}^{-}$ and $C_{m}^{+}$ are the global coordinates of the nodes at the lower and upper surfaces, respectively. $Q_{m}^{-}$ and $Q_{m}^{+}$ are the nodal displacements of the lower and upper surfaces and subscript $m$ runs from 1 to the number of degrees of freedom of each of the surfaces of the cohesive element.

The last term in Equation (\ref{eq:dDedN}), $\textrm{d}a/\textrm{d}N$, is the crack growth rate, which can be expressed by any variant of the Paris' law \cite{Paris1961} based on the energy release rate, $\mathcal{G}$. Thus, any experimentally observed effect on the crack growth rate, such as the effect of the mode mixity, $\Phi$, or the load ratio, $R$, can be included in the fatigue damage model. In this work, the crack growth rate, $\textrm{d}a$/$\textrm{d}N=f(\mathcal{G}_{max}, R,\Phi)$, is assumed to depend on the maximum cyclic energy release rate, $\mathcal{G}_{max}$, the load ratio, $R$, defined as:

\begin{equation}
R=\sqrt{\frac{\mathcal{G}_{min}}{\mathcal{G}_{max}} } 
\label{eq:R}
\end{equation}

\noindent and the energy-based mode mixity, $\Phi$:

\begin{equation}
\Phi=\frac{\mathcal{G}_{s}}{\mathcal{G}_{I}+\mathcal{G}_{s}}  
\label{eq:MM}
\end{equation}

\noindent where subscript $s$ denotes the shear component of the energy release rate. 

The following Paris' law-based expression is proposed for the crack growth rate, $\textrm{d}a$/$\textrm{d}N$, though any other expression that incorporates, for instance, the energy release threshold \cite{Brunner2017,Jones2018,Jones2017, Yao2018} could be used. For a given load ratio, $R$, the Paris' law-like expression reads:

\begin{equation}
\frac{\textrm{d} a}{\textrm{d} N}=
\left \lbrace 
\begin{split}
& A \left(\frac{ \mathcal{G}_{max}\left(1-R \right)}{\mathcal{G}_{c}} \right) ^{p}  \qquad \text{for}  \quad \mathcal{G}_{th} < \mathcal{G}_{max} < \mathcal{G}_{c}
\\& 0  \qquad \qquad \qquad  \qquad  \qquad \ \text{for}  \quad  \mathcal{G}_{max} \leq \mathcal{G}_{th}
\end{split} \right. 
\label{eq:Paris} 
\end{equation}

\noindent where the exponent, $p$, and coefficient, $A$, are mode-dependent parameters determined by \cite{Blanco2004}:

\begin{equation}
\begin{split}
p&=\Phi^{2}\left( p_{s}-p_{I}-p_{m}\right) + \Phi p_{m} + p_{I}
\\ \text{log} \left(A\right) &= \Phi^{2} \text{log} \left( \frac{A_{s}}{A_{m}A_{I}}\right) + \Phi \text{log} \left( A_{m}\right) + \text{log} \left( A_{I}\right)
\end{split}  
\label{eq:Blanco} 
\end{equation}

$p_{I}$ and $A_{I}$ are the parameters for pure mode I,  $p_{s}$ and $A_{s}$ are the parameters for shear mode, and $p_{m}$ and $A_{m}$ are mode interpolation parameters. Note that, in the original formulation presented in \cite{Blanco2004}, the pure mode II parameters, $p_{II}$ and $A_{II}$, where used instead of $p_{s}$ and $A_{s}$. Likewise in the quasi-static model \cite{Turon2006,Turon2010}, modes II and III are not disaggregated in the current implementation ($p_{s}=p_{II}=p_{III}$ and $A_{s}=A_{II}=A_{III}$). $\mathcal{G}_{th}$ in Equation (\ref{eq:Paris}) is the energy release rate threshold below which no propagation occurs. Its dependence with the mode mixity is assumed to follow a Benzeggagh-Kenane based \cite{Benzeggagh1996,Turon2007} expression:

\begin{equation}
\mathcal{G}_{th}=\mathcal{G}_{Ith}+\left(\mathcal{G}_{sth}-\mathcal{G}_{Ith} \right)\Phi^{\eta_{2}}
\label{eq:BK2} 
\end{equation}

\noindent where subscripts $Ith$ and $sth$ denote the pure mode I and shear threshold values, respectively, and $\eta2$ is an experimentally determined mode interaction parameter. 

The maximum cyclic energy release rate, $\mathcal{G}_{max}$, and the energy-based mode mixity, $\Phi$, are computed using the 3D CZ $J$-integral formulation presented in \cite{Carreras_Jint}. According to \cite{Carreras_Jint}, the integration paths cross the CZ following the GDD trajectory (c.f. Figure \ref{fig:sketch_GDD}). The computation of the $J$-integral is done using the current traction-displacement jump field over the entire integration path. Thus, any variation in the cohesive quantities due to an abrupt change in the loading scenario or geometry is captured at the current time. Each point within the cohesive zone belongs to a single integration path; and each integration path is understood as a crack propagating in the GDD at a velocity of $\textrm{d}a/\textrm{d}N$. In other words, all the integration points in the cohesive zone along a certain GDD are attributed the same $\textrm{d}a/\textrm{d}N$, which is a scalar measure  described by a Paris' law-like expression (Equation \ref{eq:Paris}). The model enforces that the speed at which the cohesive zone advances with the number of cycles is equal to  $\textrm{d}a/\textrm{d}N$.

Using the mode-decomposed 3D CZ $J$-integral formulation, the maximum cyclic energy release rate, $\mathcal{G}_{max}$, is computed as the sum of the $J$-terms:

\begin{equation}
\mathcal{G}_{max}={J}_{I}+{J}_{II}+{J}_{III}  
\label{eq:G_max2}
\end{equation}

\noindent and the energy-based mode mixity, $\Phi$, defined in Equation (\ref{eq:MM}) is evaluated as:

\begin{equation}
\Phi=\frac{{J}_{II}+{J}_{III}}{{J}_{I}+{J}_{II}+{J}_{III}}  
\label{eq:MM_3D}
\end{equation}

Note the difference between the displacement jump-based mode mixity, $B$, expressed in Equation (\ref{eq:B}) as a local quantity, and the energy release rate-based mode mixity, $\Phi$, which is extracted from the mode-decomposition of the $J$-integral. 

Finally, for the cycle jump strategy, a target for the maximum increment in crack length per solution sub-step,  $\Delta a_{t}$, is set. Based on this, the increment in cycles is determined as:

\begin{equation}
\Delta N=\frac{\Delta a_{t}}{\left( \frac{\text{d} a}{\text{d} N }\right)_{max} }
\label{eq:DN_3D}
\end{equation}

\noindent where $\left( \frac{\text{d} a}{\text{d} N }\right)_{max}$ is the instantaneous maximum crack growth rate in the model.

\begin{figure}[h!]
	\centering
	\includegraphics[width=16 cm]{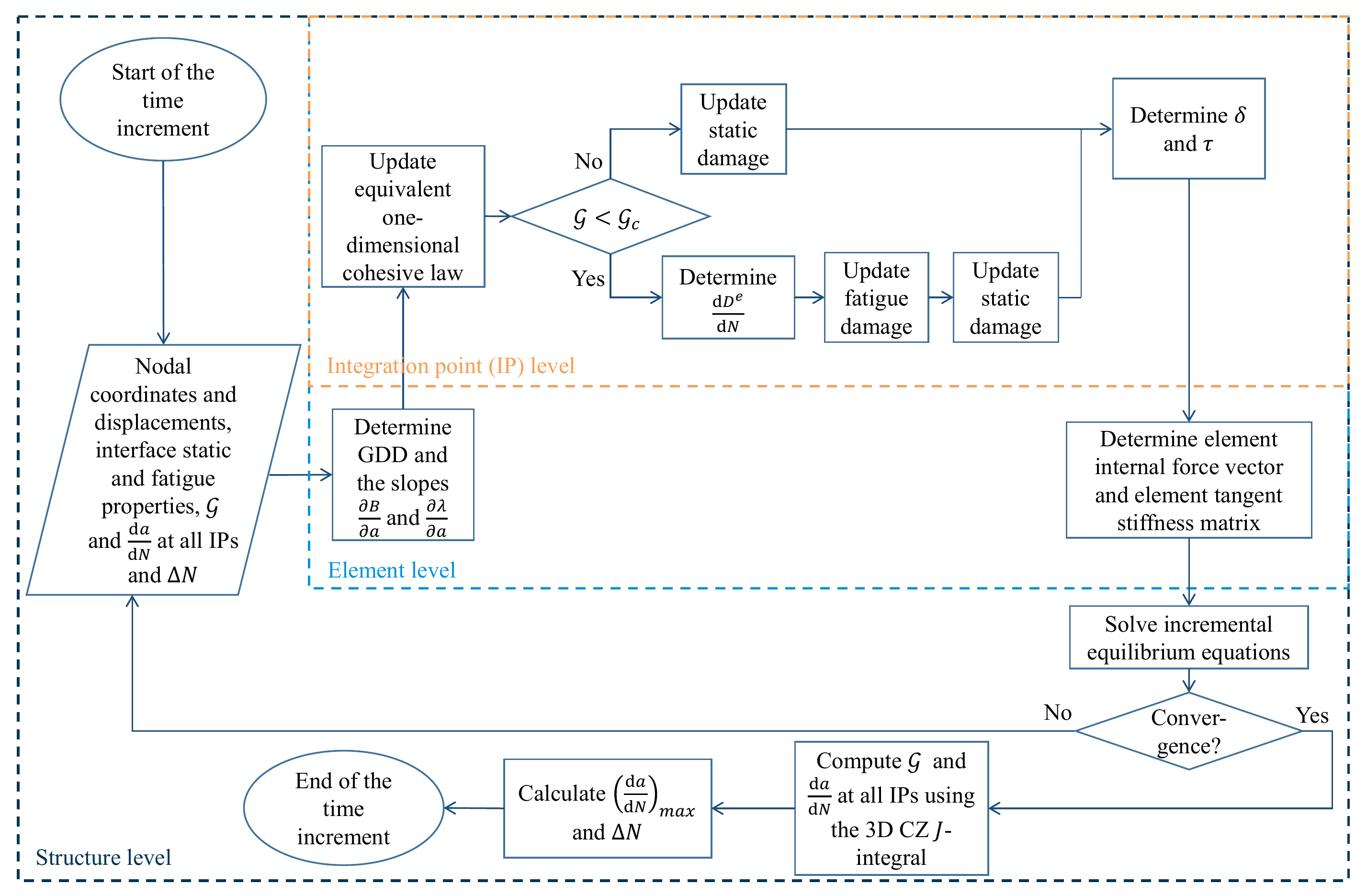} 	
	\caption{Simplified flowchart of the calculations for one time increment of the proposed simulation method for fatigue-driven delamination.}
	\label{fig:flowchart}	
\end{figure}

The formulation described in this section has been implemented as a user-defined cohesive element in the commercial finite element code Abaqus. The implemented procedure for the calculations of the simulation method for fatigue-driven delamination is summarized in the flowchart shown in Figure \ref{fig:flowchart}.

\section{Comparison with experimental results} \label{sec:verification}

In order to evaluate the performance of the presented method, two different studies are performed. First, the capability of accurately reproduce the phenomenological expression used as input in the model for the crack growth rate is assessed in Section \ref{sec:val2D}: \textit{Verification}. This is done through 2D simulations of the same specimens that provided the Paris' law-based expression under different mode mixities, $\Phi$, and load ratios, $R$. Secondly, the  capability of the method to accurately predict delamination growth in problems that cannot be simplified to 2D models is assessed in Section \ref{sec:val3D}: \textit{Validation}. The benchmark case presented in \cite{demo_exp,Carreras_data}, which exhibits change in crack front shape and growth rate, is used for this purpose.

All the specimens used in these studies were made of unidirectional (UD) Carbon Fibre Reinforced Polymer (CFRP) prepreg plies of 0.184 mm nominal thickness stacked at 0$^\circ$. AERNNOVA Engineering, the company supplying the material, provided the elastic properties for the unidirectional laminate material (Table \ref{tab:elast_prop}). The static interlaminar fracture properties are listed in Table \ref{tab:props_model_values_static}. The interlaminar fracture toughness were measured following the standards recommendations \cite{ASTM_modeI,ASTM_mmode,ASTM_modeII} under 4 different mode mixity conditions: 0\% ($\mathcal{G}_{Ic}$), 50\%, 75\% and 100\% ($\mathcal{G}_{IIc}$). The nominal values are presented in \cite{demo_exp}. The Benzeggagh-Kenane \cite{Benzeggagh1996} mode interpolation parameter, $\eta$, in Equation (\ref{eq:BK}) has been fitted to the results and included in Table \ref{tab:props_model_values_static}. Moreover, the interfacial shear strength, $\tau_{IIo}$, was measured following the ASTMD2344 standard. The obtained nominal value was 98 MPa. Then, the interfacial tensile strength, $\tau_{Io}$, is expressed as a function of $\tau_{so}$, $\mathcal{G}_{Ic}$ and $\mathcal{G}_{sc}$ (c.f. Equation (\ref{eq:tauIo_tauIIo})) \cite{Turon2010}:

\begin{equation}
\tau_{Io}=\tau_{so}\sqrt{\frac{\mathcal{G}_{Ic}}{\mathcal{G}_{sc}}}
\label{eq:tauIIo_tauIo} 
\end{equation}

\begin{table}[h!]
	\centering
	
	\begin{tabular}{lrlllrl}
		\hline \hline
		\multicolumn{3}{c}{Laminate properties} \\
		\hline 
		$E_{11}$: Longitudinal Young's modulus & 154& GPa \\
		$E_{22}=E_{33}$: Transversal Young's modulus & 8.5& GPa \\
		$G_{12}=G_{13}$: Shear modulus in the longitudinal planes& 4.2& GPa \\
		$G_{23}$: Shear modulus in the transversal plane& 3.04& GPa \\
		$\mu_{12}=\mu_{13}$: Poisson's coefficient  in the longitudinal planes & 0.35& \multicolumn{1}{c}{-} \\
		$\mu_{23}$: Poisson's coefficient  in the transversal plane & 0.4& \multicolumn{1}{c}{-}\\
		\hline \hline      
	\end{tabular}
	\caption{Laminate elastic properties of the validation material. }	
	\label{tab:elast_prop}
\end{table}

\begin{table}[h!]
	\centering
	
	\begin{tabular}{lrl}
		\hline \hline
		Property          &  Value   & Units\\ \hline 
		$\mathcal{G}_{Ic}$ &  0.305 & N/mm\\
		$\mathcal{G}_{IIc}$ & 2.77  & N/mm\\
		$\eta$ &  2.05&\multicolumn{1}{c}{-} \\
		$\tau_{Io}$ &  32.6 & MPa \\
		$\tau_{IIo}$ &  98 & MPa \\					 				
		\hline \hline    
	\end{tabular}
	\caption{Static interlaminar fracture properties of the validation material.}	
	\label{tab:props_model_values_static}
\end{table}

\subsection{Verification} \label{sec:val2D}

Crack growth rate curves under mode I loading conditions were obtained using the multi-specimen testing methodology described in \cite{Multi} with $R=0.1$ and $R=0.5$. Under mode II and mixed-mode loading conditions, the methodology described in \cite{Carreras2017} and \cite{Jaeck2018} was used instead. This consisted of varying the applied cyclic displacement in order to increase the energy release rate range covered by a single test. Two different load ratios ($R=0.1$ and $R=0.3$) were tested under mode II conditions and two different mode mixities ($\Phi=0.5$ and $\Phi=0.75$) were tested with $R=0.1$. The Paris' law-based fitting parameters (according to Equation (\ref{eq:Paris})) from all the mentioned crack growth rate curves are listed in Table \ref{tab:props_models_2D}.

\begin{table}[h!]
	\centering
	
	\begin{tabular}{lrrl}
		\hline \hline
				Property          & & & Units\\ \hline
		\multicolumn{4}{c}{Mode I $\Phi$=0} \\
		
		&$R=0.1$ &$R=0.5$& \\	\\	 		
		$p$ &8.39  & 13.8& \multicolumn{1}{c}{-} \\
		$A$ & 6.45E-2& 8.26E2 & mm/cycle \\	
		\hline	 	
		\multicolumn{4}{c}{Mode II $\Phi$=1} \\
	
		&$R=0.1$ &$R=0.3$& \\		\\		
		$p$ &3.62  &4.17 & \multicolumn{1}{c}{-} \\
		$A$ & 7.03E-1& 7.41E-1 & mm/cycle \\ 		
			\hline	 	
		\multicolumn{4}{c}{Mixed-mode $R$=0.1} \\

		&$\Phi=0.5$ &$\Phi=0.75$& \\		 \\		
		$p$ &5.95 &4.14 & \multicolumn{1}{c}{-} \\
		$A$ & 1.26&  3.96E-1& mm/cycle \\ 		
		\hline \hline    
	\end{tabular}
	\caption{Paris' law-based fitting parameters (Equation (\ref{eq:Paris})) used in 2D simulations.}	
	\label{tab:props_models_2D}
\end{table}

\begin{figure}[h!]
	\centering
	\includegraphics[width=8 cm]{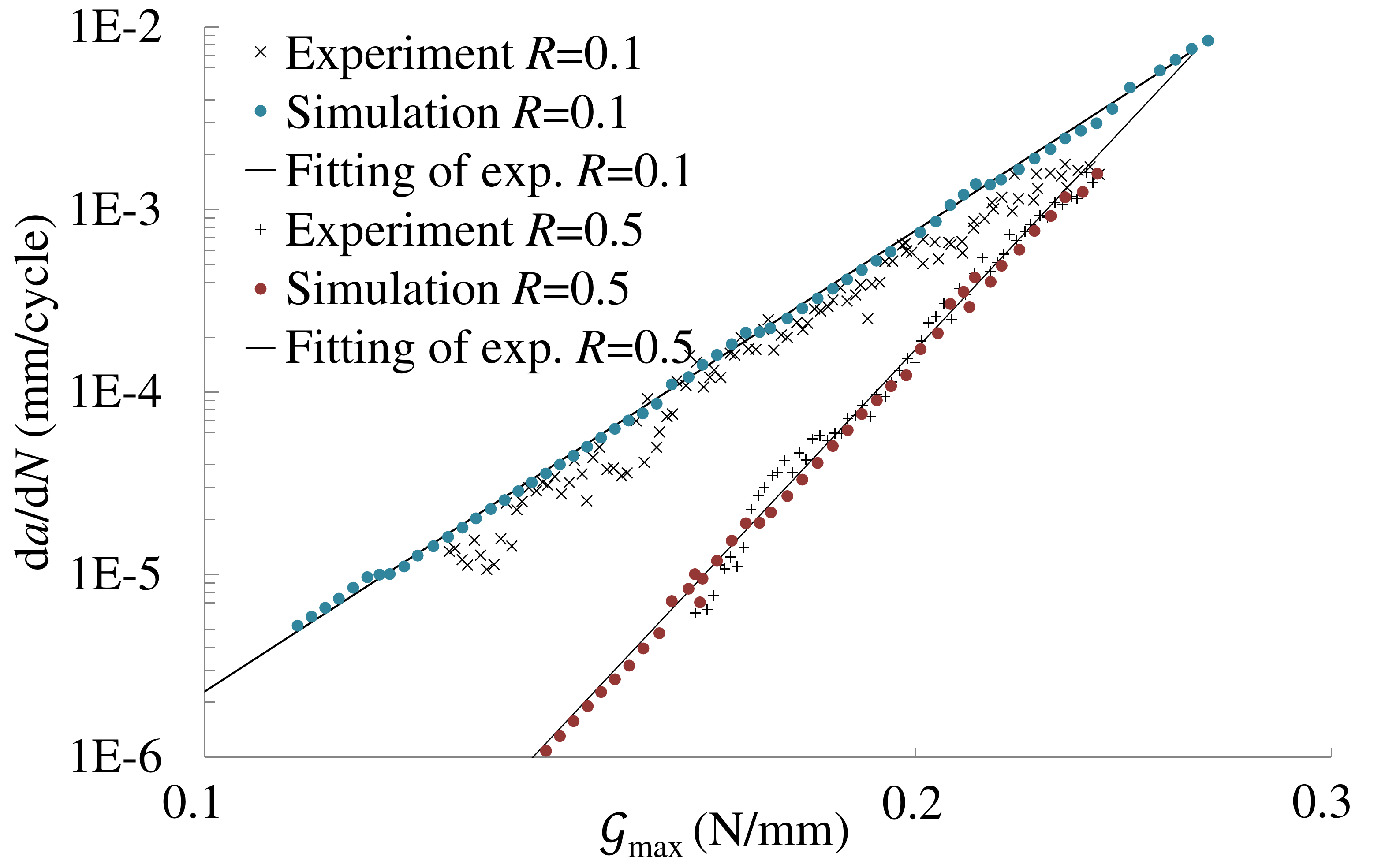} 	
	\caption{Comparison between the experimentally obtained mode I crack growth rate curves \cite{Multi} and the results from the 2D simulations. The input parameters $p_{I}$ and $A_{I}$ are obtained from the Paris' law-based fitting of the experimental data listed in Table \ref{tab:props_models_2D}.}
	\label{fig:Paris_modeI}	
\end{figure}

\begin{figure}[h!]
	\centering
	\includegraphics[width=8 cm]{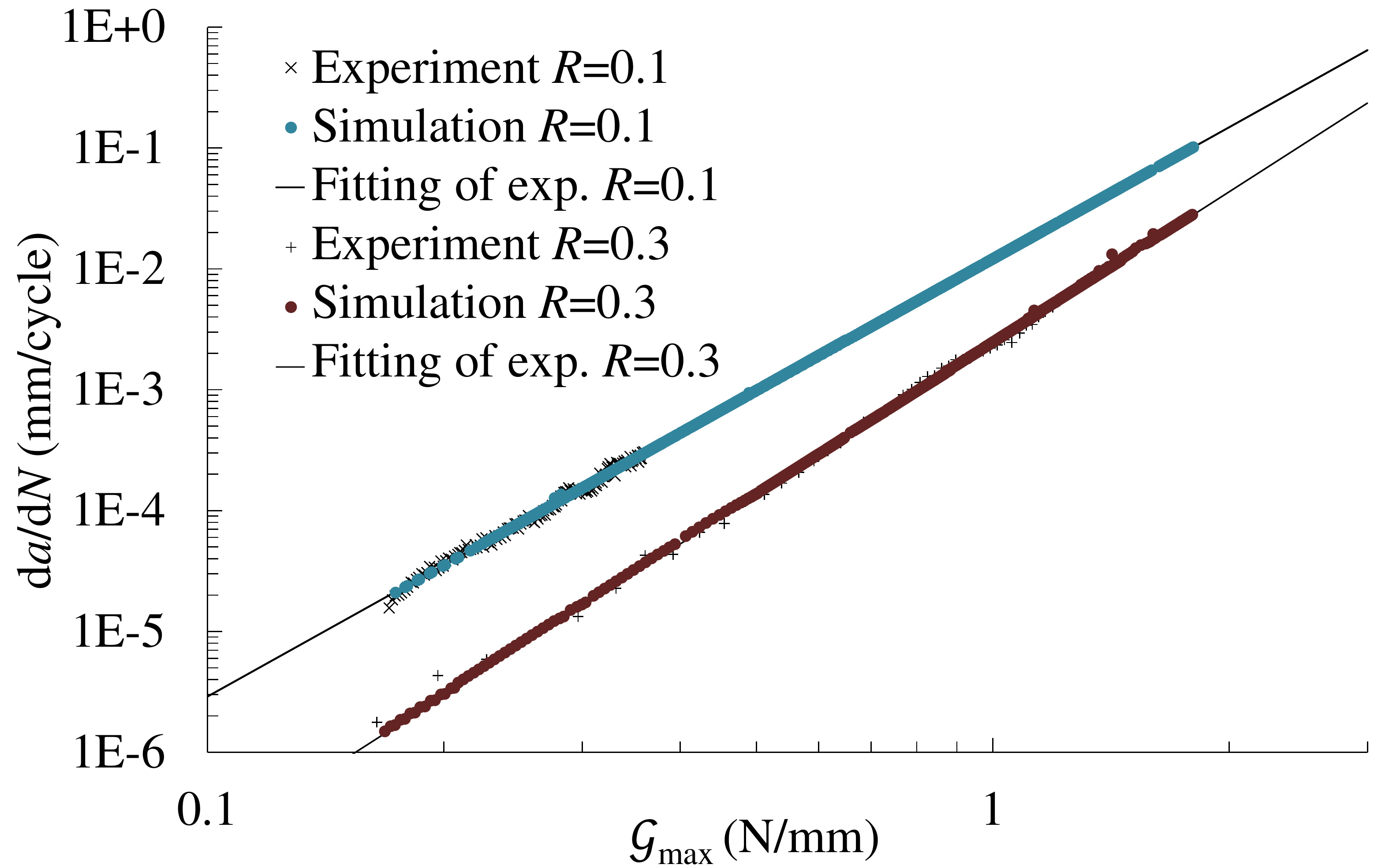} 	
	\caption{Comparison between the experimentally obtained mode II crack growth rate curves \cite{Carreras2017} and the results from the 2D simulations. The input parameters $p_{II}$ and $A_{II}$ are obtained from the Paris' law-based fitting of the experimental data listed in Table \ref{tab:props_models_2D}.}
	\label{fig:Paris_modeII}	
\end{figure}

\begin{figure}[h!]
	\centering
	\includegraphics[width=8 cm]{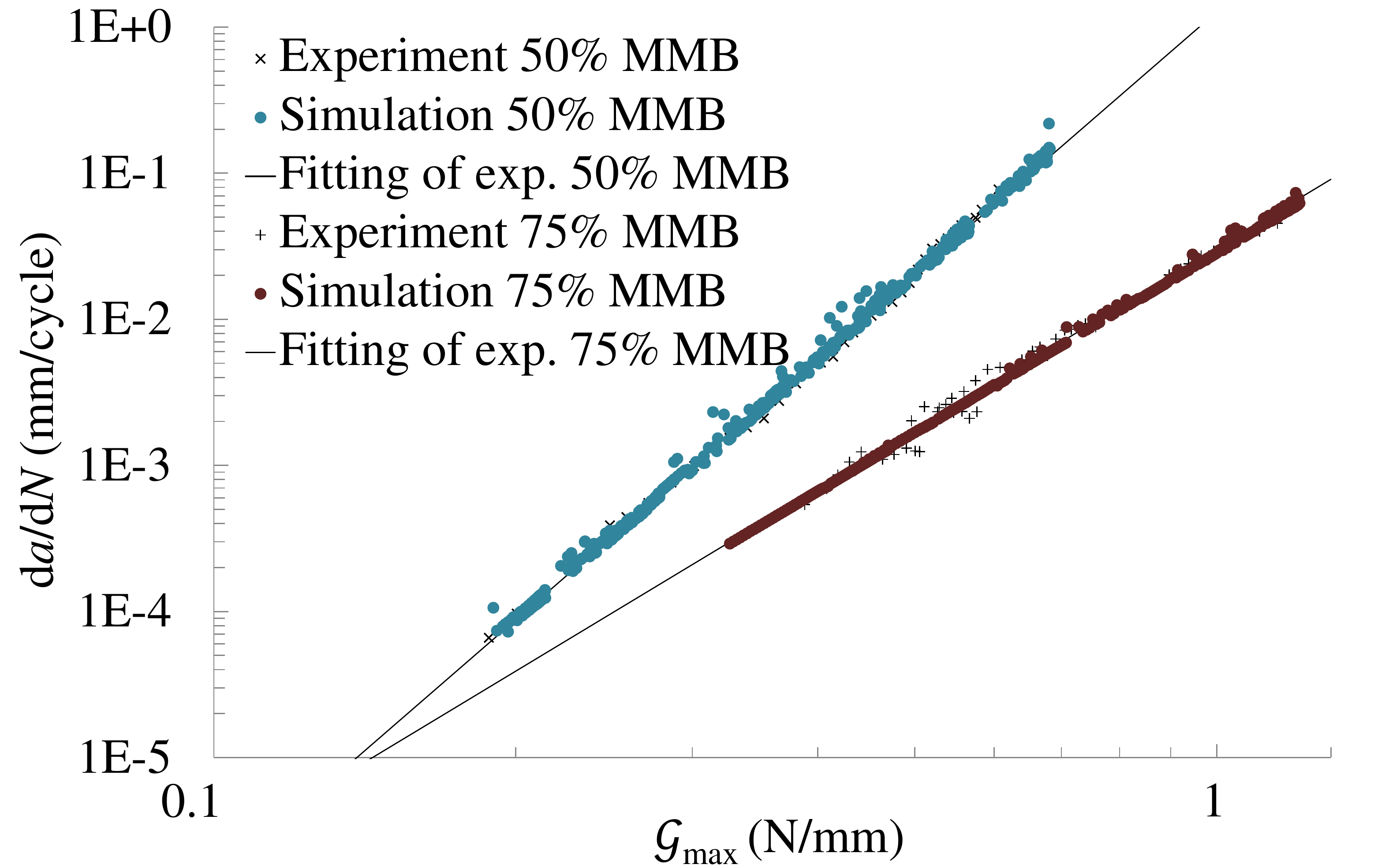} 	
	\caption{Comparison between the experimentally obtained 50\% and 75\% mixed-mode crack growth rate curves with load ratio $R=0.1$ \cite{Jaeck2018} and the results from the 2D simulations. The input parameters $p_{50\%}$, $A_{50\%}$, $p_{75\%}$ and $A_{75\%}$ are obtained from the Paris' law-based fitting of the experimental data listed in Table \ref{tab:props_models_2D}.}
	\label{fig:Paris_modeMM}	
\end{figure}

To validate the capability of the model to reproduce the phenomenological expressions used as inputs, these tests have been simulated with 2D FE models assuming plane strain conditions. The specimens were made of 16 layers of unidirectional CFRP and were 25 mm wide. The details of the FE models and analyses are given in Table \ref{tab:prop_analysis2D}. Note that due to fine mesh discretization, a standard 2x2 Newton-Cotes quadrature rule has been used to numerically integrate element quantities as opposed to more advanced integration schemes \cite{Bak2014,lindgaard2017user}, which are needed for coarser meshes. The results from the simulations are shown in figures \ref{fig:Paris_modeI} (mode I), \ref{fig:Paris_modeII} (mode II) and \ref{fig:Paris_modeMM} (mixed-mode) together with the experimental data and the fitted Paris' law curves. The crack growth rate in the simulations is determined at every converged increment as:

\begin{equation}
\frac{\textrm{d} a}{\textrm{d} N} \approx \frac{\Delta n_{\mathcal{D}} L_{e}}{\Delta N}
\label{eq:deltaa_deltaN} 
\end{equation}

\noindent where $\Delta n_{\mathcal{D}}$ is the increment in number of damaged elements, $L_{e}$ is the element length and $\Delta N$ is the increment in number of cycles. The results from the simulations accurately reproduce the crack growth rate obtained from experimental tests for all the mode mixities, $\Phi$, and load ratios, $R$, analysed.

Note that the threshold region is not explored in the experimental tests. Therefore, the energy release rate threshold is set to $\mathcal{G}_{th}=0$  in the simulations without influencing the results in figures \ref{fig:Paris_modeI} (mode I), \ref{fig:Paris_modeII} (mode II) and \ref{fig:Paris_modeMM} (mixed-mode).

\begin{table}[h!]
	\centering
	
	\begin{tabular}{lc}
		\hline \hline
		Solver    &     Newton-Raphson\\ 
		& Linear static \\
		
		Increment & Adaptive increment size	\\	 		
   
	  Element type in beams& C3D8 (Abaqus 6.12)	\\	
	  Elements in beam thickness & 4 	\\
	  Integration of cohesive element & 2 x 2 Newton-Cotes quadrature \\
	   $L_{e}$: Cohesive element length & 0.2 mm	\\
	  	  $\Delta a_{t}$: Crack increment target & 0.2 mm	\\	
	  	  $K$: Penalty stiffness & 1E5 N/mm$^3$ \\	
	  	  ACI \cite{Bak2016}: & On \\
		\hline \hline    
	\end{tabular}
	\caption{Analysis and model settings of 2D simulations. }	
	\label{tab:prop_analysis2D}
\end{table}

\subsection{Validation} \label{sec:val3D}

The capabilities of the simulation method for predicting propagation of fatigue-driven delamination are validated against the experimental benchmark case presented in \cite{demo_exp,Carreras_data}. The test specimen is a double cantilever beam (DCB) built by stacking 16 unidirectional CFRP plies at 0$^\circ$ with an initial straight delamination front made by a Teflon insert at the midplane. Two reinforcements, made of 8 plies of the same material and orientation, are bonded on top and bottom sides of the specimen in order to promote curved delamination front. The benchmark case generates a rich phenomenology of crack advance in terms of varying crack growth rate and crack front shape.

\begin{figure}[h!]
	\centering
	\includegraphics[width=10 cm]{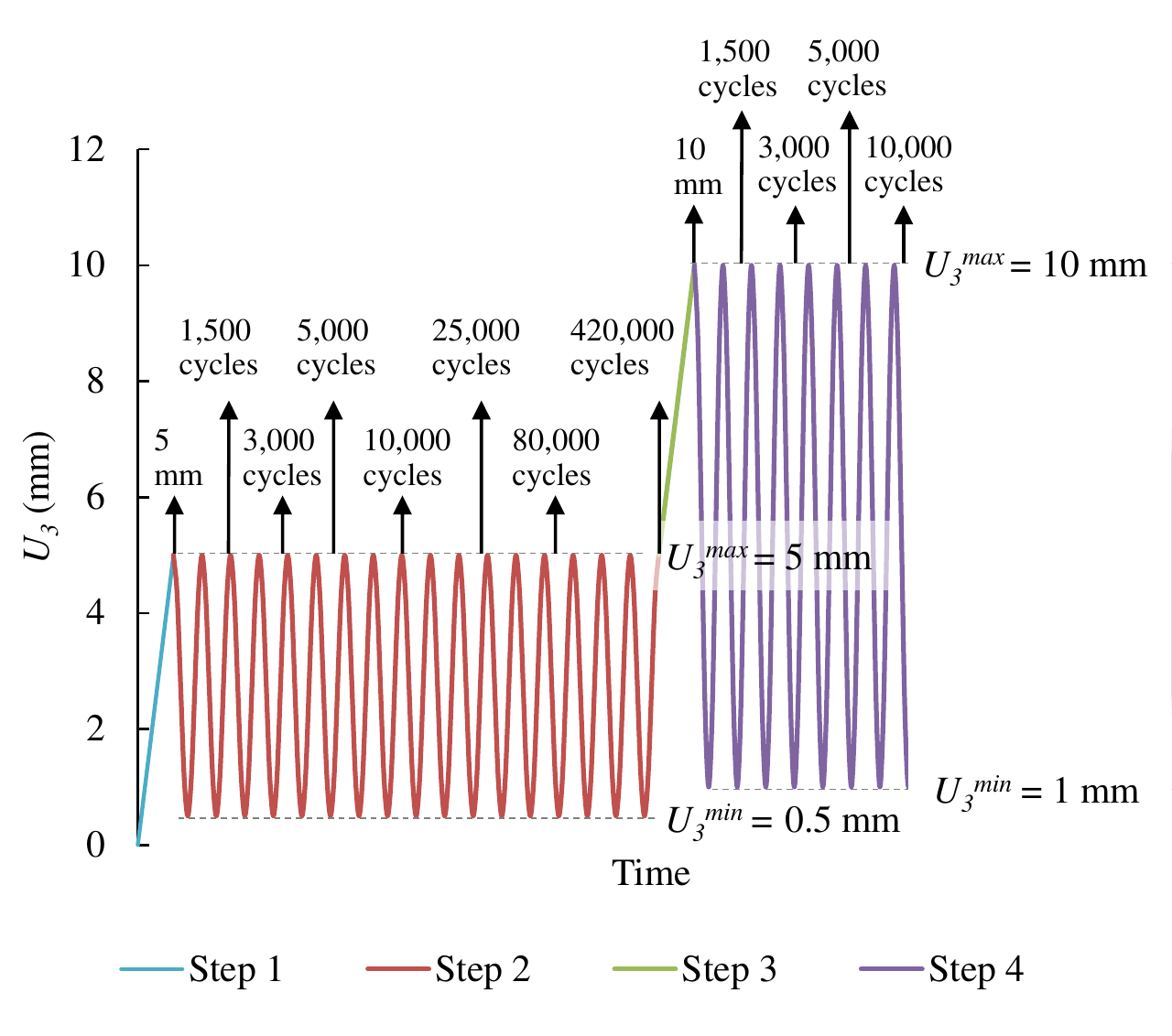} 	
	\caption{Loading steps of the benchmark test and programmed stops to monitor the delamination front.}
	\label{fig:App_displ}	
\end{figure}

The test consisted on four loading steps (c.f. Figure \ref{fig:App_displ}): 

\begin{itemize}
\item 1) Quasi-static step from the initial unloaded position until 5 mm of the prescribed displacement at a loading rate of 1 mm/min. 
\item 2) Fatigue step with a maximum cyclic displacement of 5 mm and $R=0.1$ during 420,000 cycles. 
\item 3) Quasi-static step until 10 mm of the prescribed displacement at a loading rate of 1mm/min. 
\item 4) Fatigue step with a maximum cyclic displacement of 10 mm and $R=0.1$ during 10,000 cycles.
\end{itemize}

The experimental tests were stopped at the programmed stops indicated with black arrows in Figure \ref{fig:App_displ} to perform X-ray inspection of the delamination front location. A batch of three specimens (numbered 1, 2 and 3) were tested following the described loading sequence. The reader is referred to \cite{demo_exp} for further details on the employed test methodology and data reduction techniques. 

\begin{figure}[h!]
	\centering
	\includegraphics[width=14 cm]{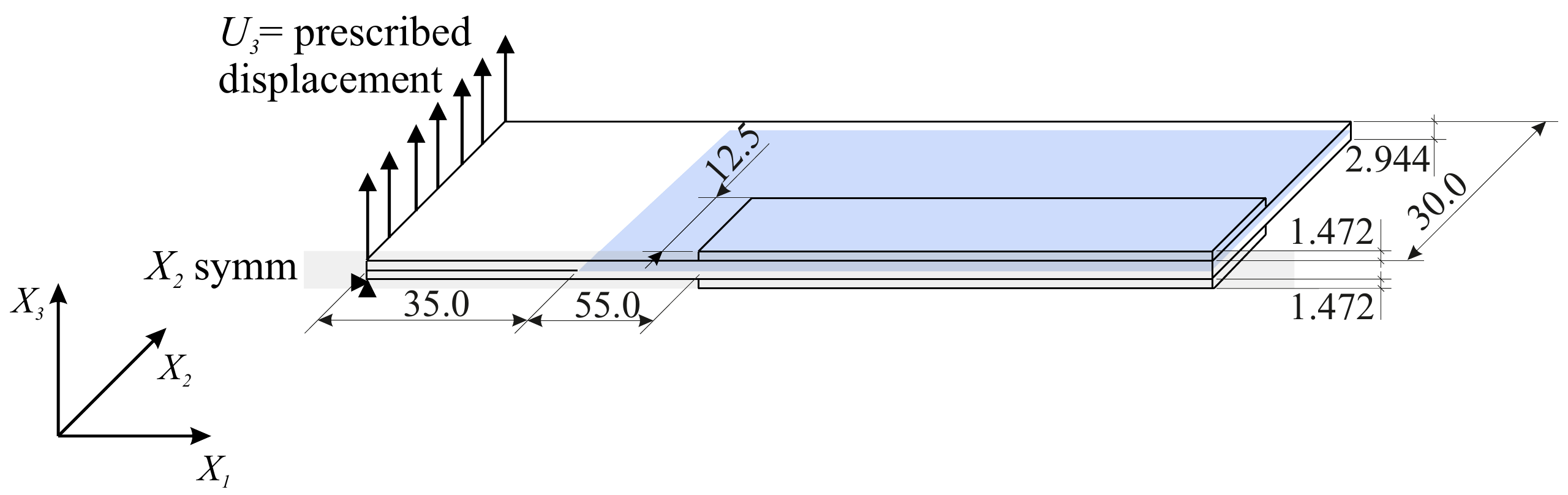} 	
	\caption{Sketch of the partially reinforced DCB specimen. Only the mid-surface area shadowed in blue is represented in figures \ref{fig:Demo_fat}, \ref{fig:Demo_fat2}, \ref{fig:Sim_resul} and \ref{fig:Sim_resul2}. Units in mm.}
	\label{fig:Demostrador_model_Tesi}	
\end{figure}

Exploiting $X_{2}$-symmetry, only one half of the specimen was modeled to reduce the required computational resources. The dimensions and boundary conditions of the partially reinforced DCB specimen model are shown in Figure \ref{fig:Demostrador_model_Tesi}. The simulation and model settings of the simulations are listed in Table \ref{tab:prop_analysis3D}. Finally, the fatigue properties used as input parameters for the user-defined cohesive elements were obtained from \cite{demo_exp} and listed in Table \ref{tab:props_model_valuesfatigue}. The $p_{m}$ and $A_{m}$ were obtained by fitting Equation (\ref{eq:Blanco}) to the $p$ and $A$ parameters for pure mode I, mode II, 50\% and 75\% mode-mixities with $R=0.1$ listed in Table \ref{tab:prop_analysis2D}. The energy release rate threshold, $\mathcal{G}_{th}$, was not mode-interpolated using Equation (\ref{eq:BK2}) because there was no mode-mixity data available that could be used to determine the mode interaction parameter, $\eta_{2}$. Instead, a linear interpolation was used. This is not affecting the results since the loading conditions are very lose to pure mode I loading. The pure mode values, $\mathcal{G}_{Ith}$ and $\mathcal{G}_{IIth}$, were obtained from \cite{demo_exp}.  

\begin{table}[h!]
	\centering
	
	\begin{tabular}{lc}
		\hline \hline
		Solver    &     Newton-Raphson\\ 
		& Linear static \\
		
		Increment & Adaptive increment size	\\	 		
		
		Element type in beams& SC8R (Abaqus 6.12)	\\	
		Elements in beam thickness & 1 	\\
		Elements in reinforcement thickness & 1 	\\	
		Integration of cohesive element & 2 x 2 Newton-Cotes quadrature \\
		$L_{X_1}$: Cohesive element length & 0.36810 mm	\\
		$L_{X_2}$: Cohesive element width & 0.21605 mm	\\
		$\Delta a_{t}$: Crack increment target & 0.36 mm	\\
		$K$: Penalty stiffness & 1E5 N/mm$^3$ \\	
			  	  ACI \cite{Bak2016}: & On \\	
		\hline \hline    
	\end{tabular}
	\caption{Analysis and model settings of 3D simulations. }	
	\label{tab:prop_analysis3D}
\end{table}

\begin{table}[h!]
	\centering
	
	\begin{tabular}{lrl}
		\hline \hline
		Property          &  Value   & Units\\ \hline 			
		$p_{I}$ &8.39 & \multicolumn{1}{c}{-}  \\
		$A_{I}$ & 6.45E-2& mm/cycle \\
		$p_{II}$  & 3.62&  \multicolumn{1}{c}{-} \\
		$A_{II}$  & 7.03E-1& mm/cycle  \\
		$p_{m}$  &-6.20 &  \multicolumn{1}{c}{-} \\
		$A_{m}$  & 1.28E5&  mm/cycle \\
		$\mathcal{G}_{Ith}$  & 8.54E-2& N/mm\\
		$\mathcal{G}_{IIth}$  & 8.29E-2 & N/mm\\		 
		$\eta_{2}$ &\multicolumn{1}{c}{\ \ \ \ \ \ \ \ -} & \multicolumn{1}{c}{-} \\	
		$R$ &  0.1  & \multicolumn{1}{c}{-} \\			 				
		\hline \hline    
	\end{tabular}
	\caption{Fatigue properties used in 3D simulations.}	
	\label{tab:props_model_valuesfatigue}
\end{table}

\begin{figure}[h!]
	\centering
	\includegraphics[width=9 cm]{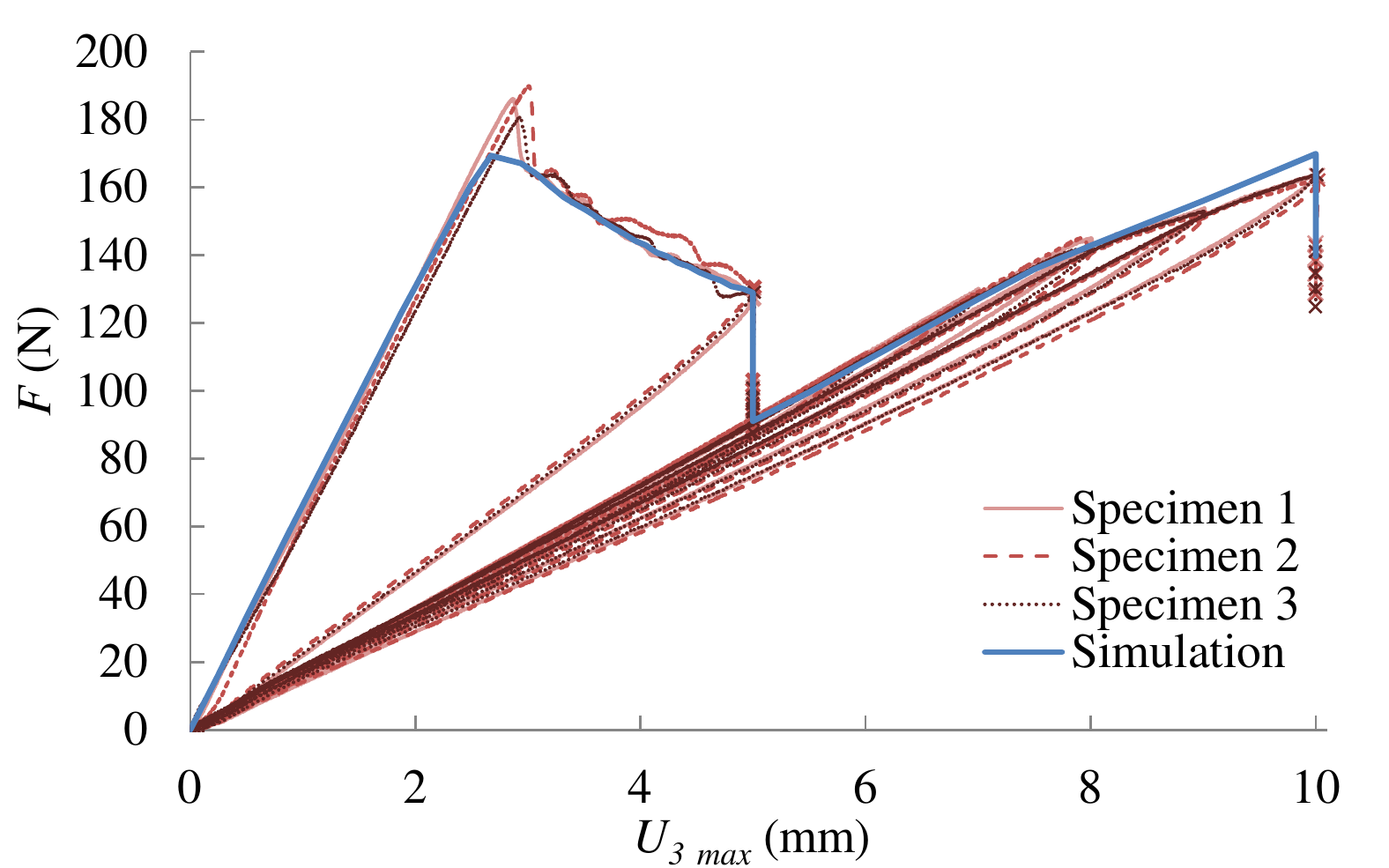} 	
	\caption{Comparison of numerical \cite{Carreras_data_num} and experimental results \cite{demo_exp,Carreras_data} for force versus displacement from fatigue testing on the demonstrator specimen. The dependence between the prescribed displacement, $U_{3}$, and the number of cycles, $N$, is shown in Figure \ref{fig:App_displ}.}
	\label{fig:F_d}	
\end{figure}

The force-displacement relationship of the 3 specimens \cite{demo_exp,Carreras_data} is plotted in Figure \ref{fig:F_d} together with the simulation results. Note that during fatigue steps 2 and 4, the maximum cyclic displacement is constant at 5mm and 10 mm, respectively. The simulation method accurately predicts the mechanical response observed in the experiments.

\begin{figure}[h]
	\centering
	\begin{tabular}{lll}
		\includegraphics[width=50mm]{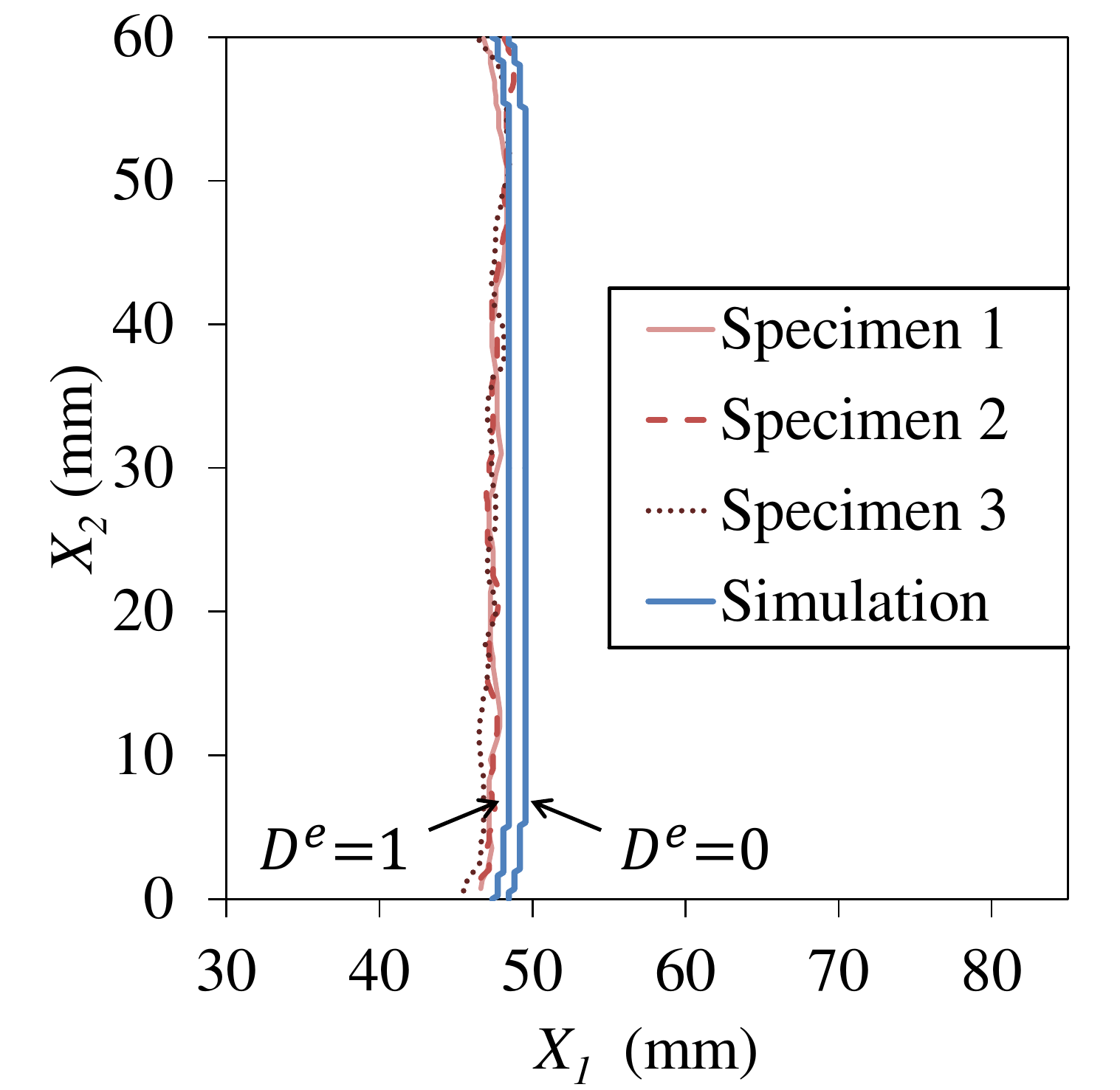} & \includegraphics[width=50mm]{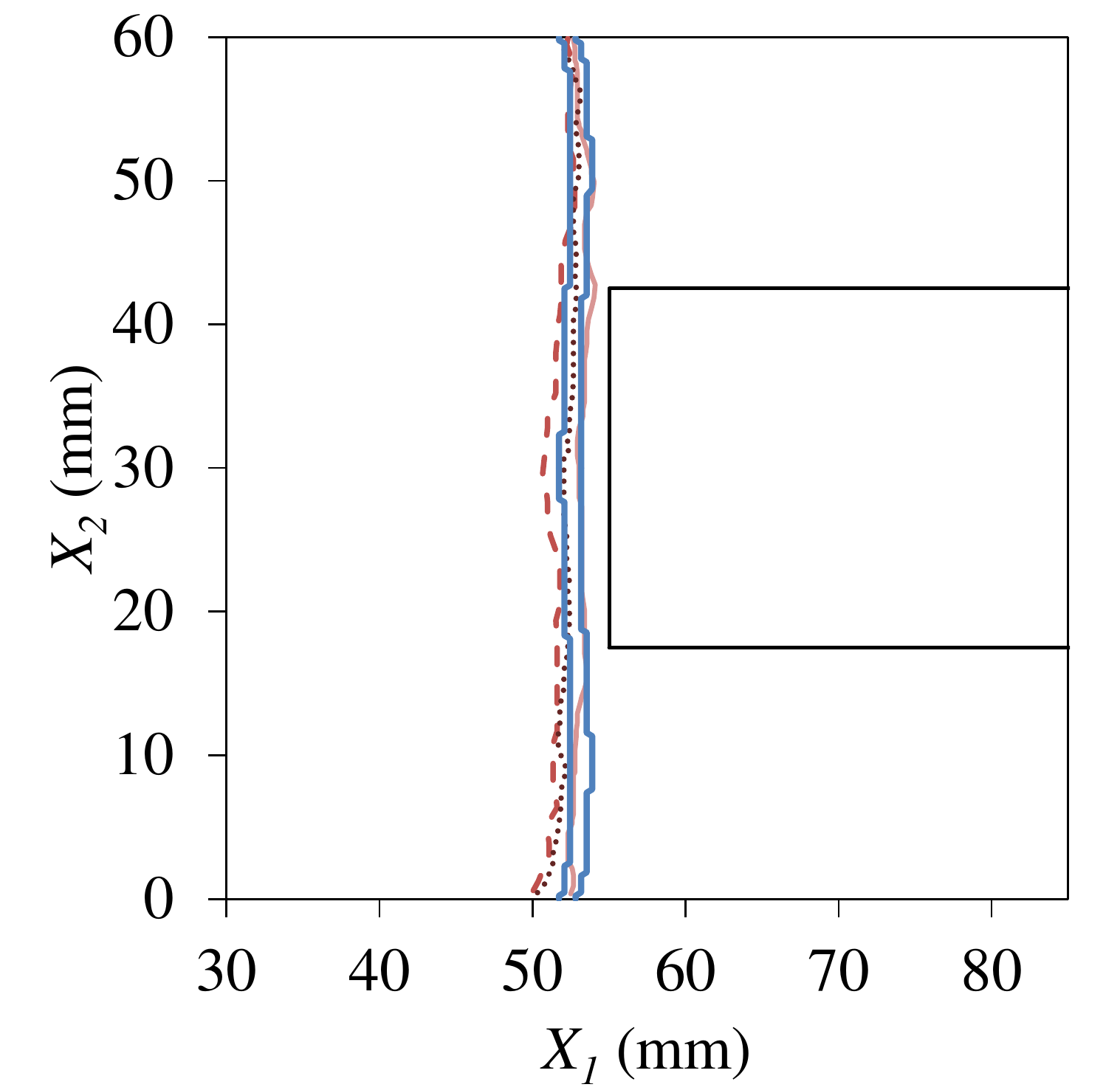}& \includegraphics[width=50mm]{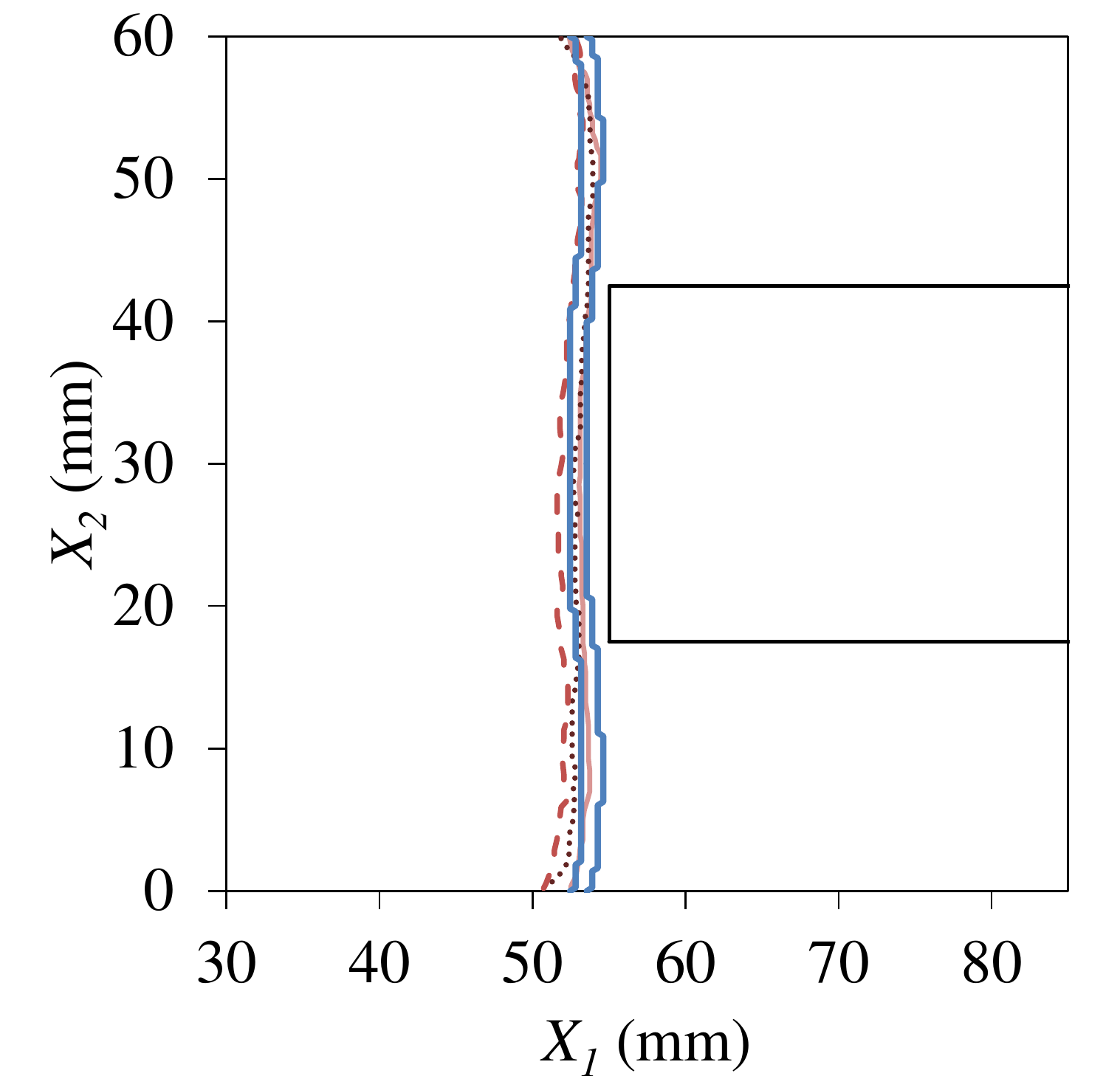} \\
		\ \ \ \ \ \ a) Step 1. Static loading & \ \ \ \ \ \ b) Step 2. Fatigue loading & \ \ \ \ \ \ c) Step 2. Fatigue loading \\
		\ \ \ \ \ \ $U_{3}= 5$ mm	& \ \ \ \ \ \ $U_{3}^{max}= 5$ mm & \ \ \ \ \ \ $U_{3}^{max}= 5$ mm \\	
		& \ \ \ \ \ \ $N= 1,500$ cycles & \ \ \ \ \ \ $N= 3,000$ cycles\\	
		\includegraphics[width=50mm]{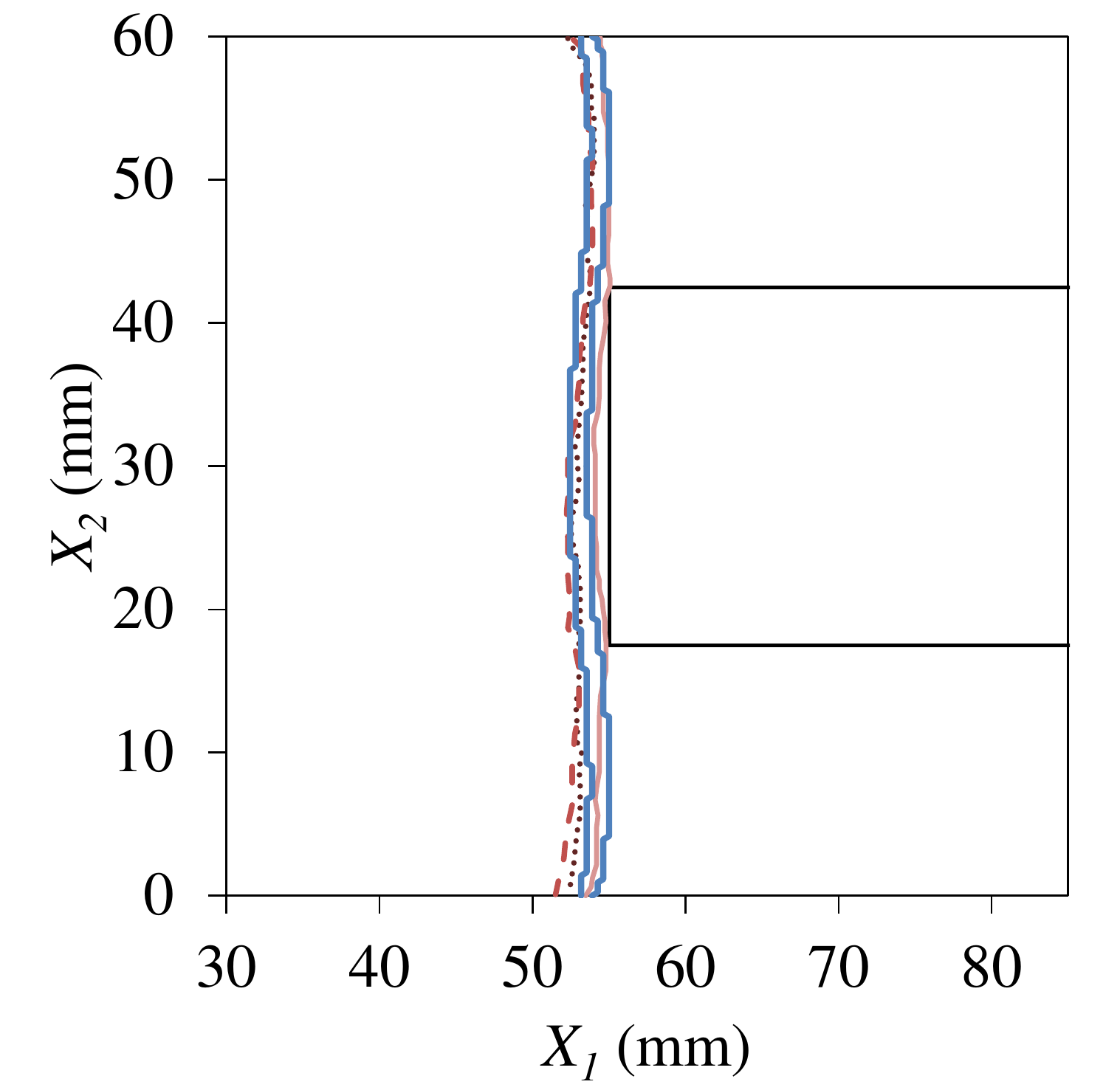} & \includegraphics[width=50mm]{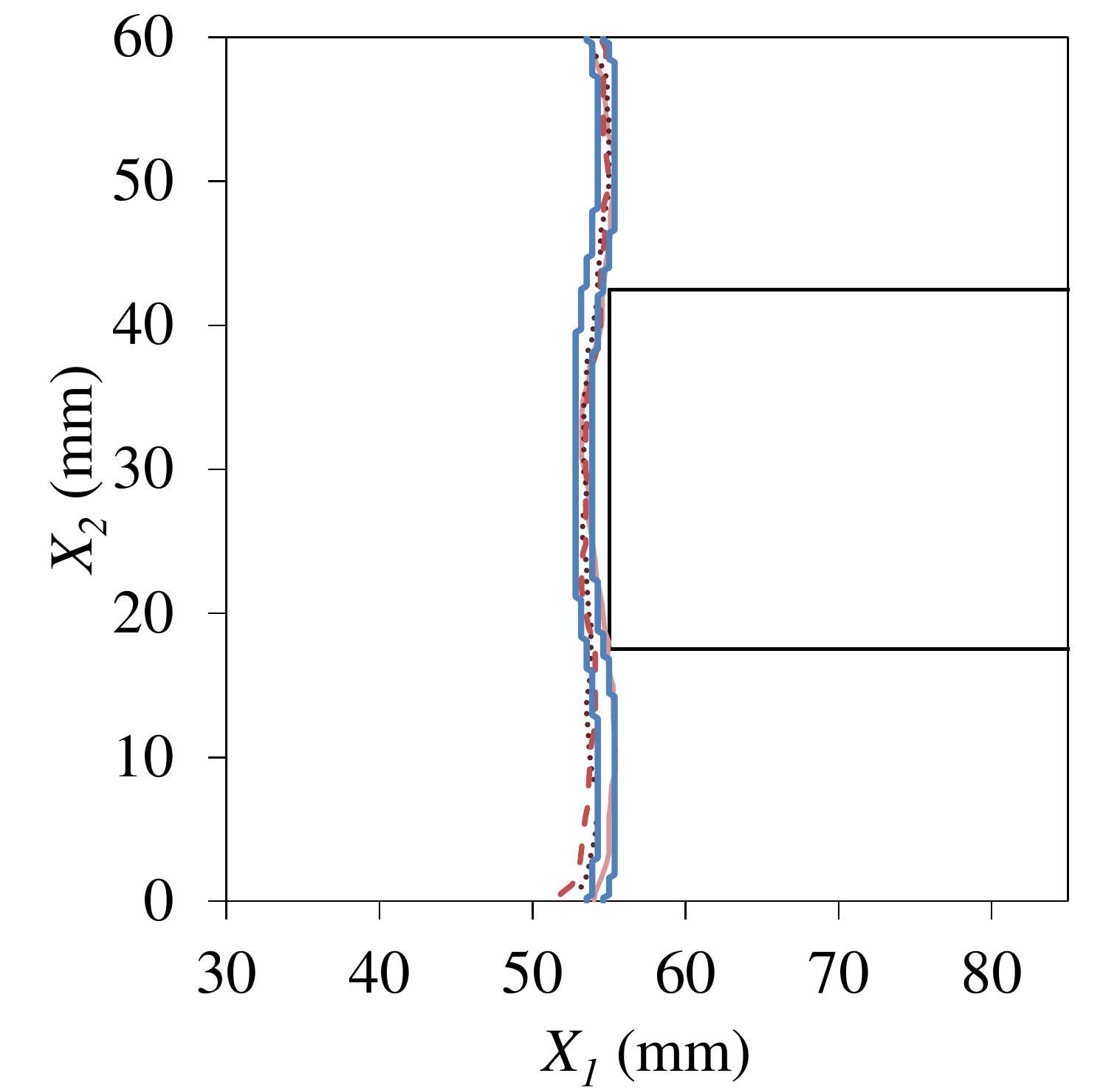} & \includegraphics[width=50mm]{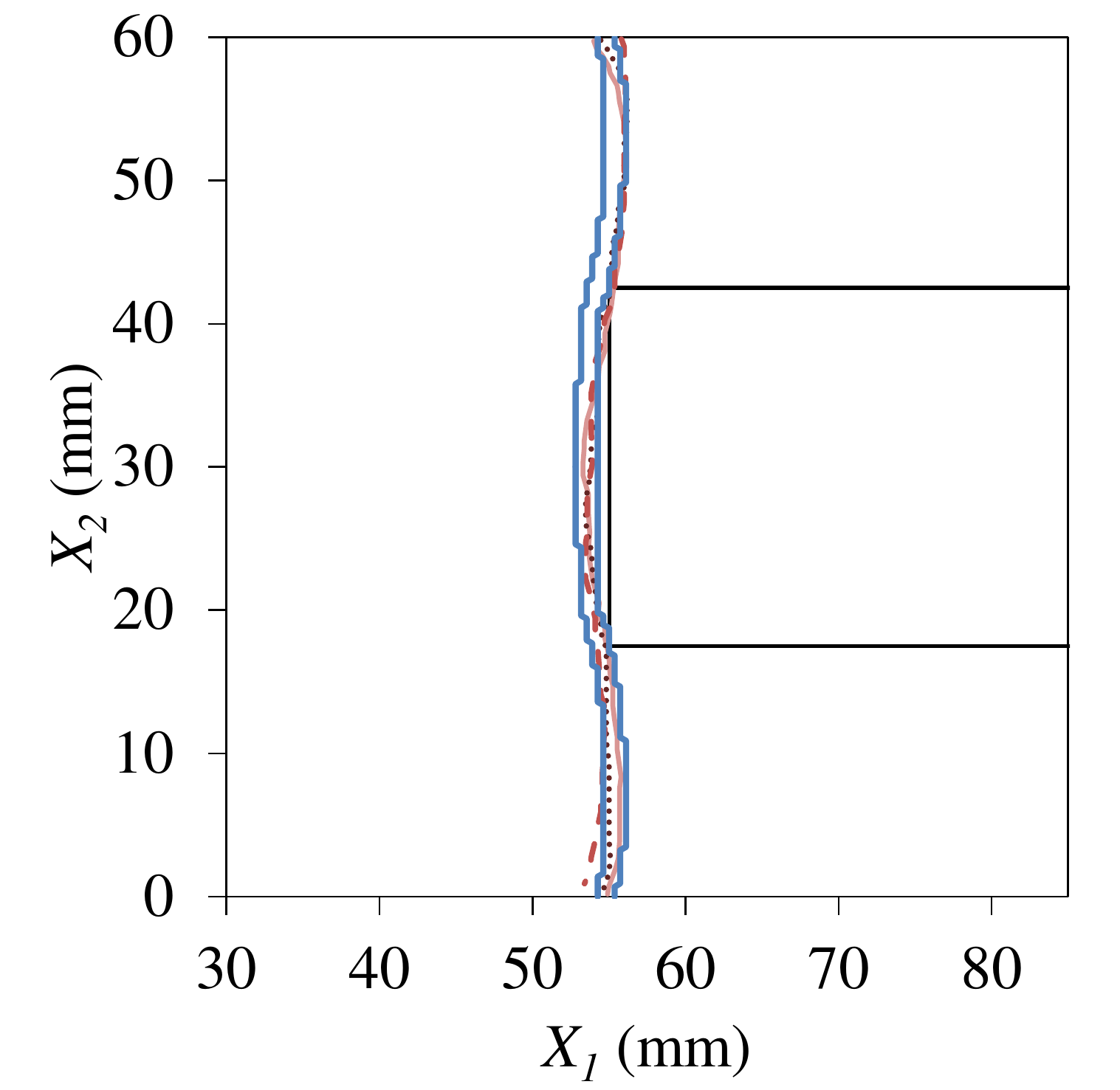}\\
		\ \ \ \ \ \ d) Step 2. Fatigue loading & \ \ \ \ \ \ e) Step 2. Fatigue loading &\ \ \ \ \ \ f) Step 2. Fatigue loading \\
		\ \ \ \ \ \  $U_{3}^{max}= 5$ mm 	& \ \ \ \ \ \ $U_{3}^{max}= 5$ mm & \ \ \ \ \ \ $U_{3}^{max}= 5$ mm  \\		
		\ \ \ \ \ \   $N= 5,000$ cycles	& \ \ \ \ \ \ $N= 10,000$ cycles & \ \ \ \ \ \ $N= 25,000$ cycles\\
		\includegraphics[width=50mm]{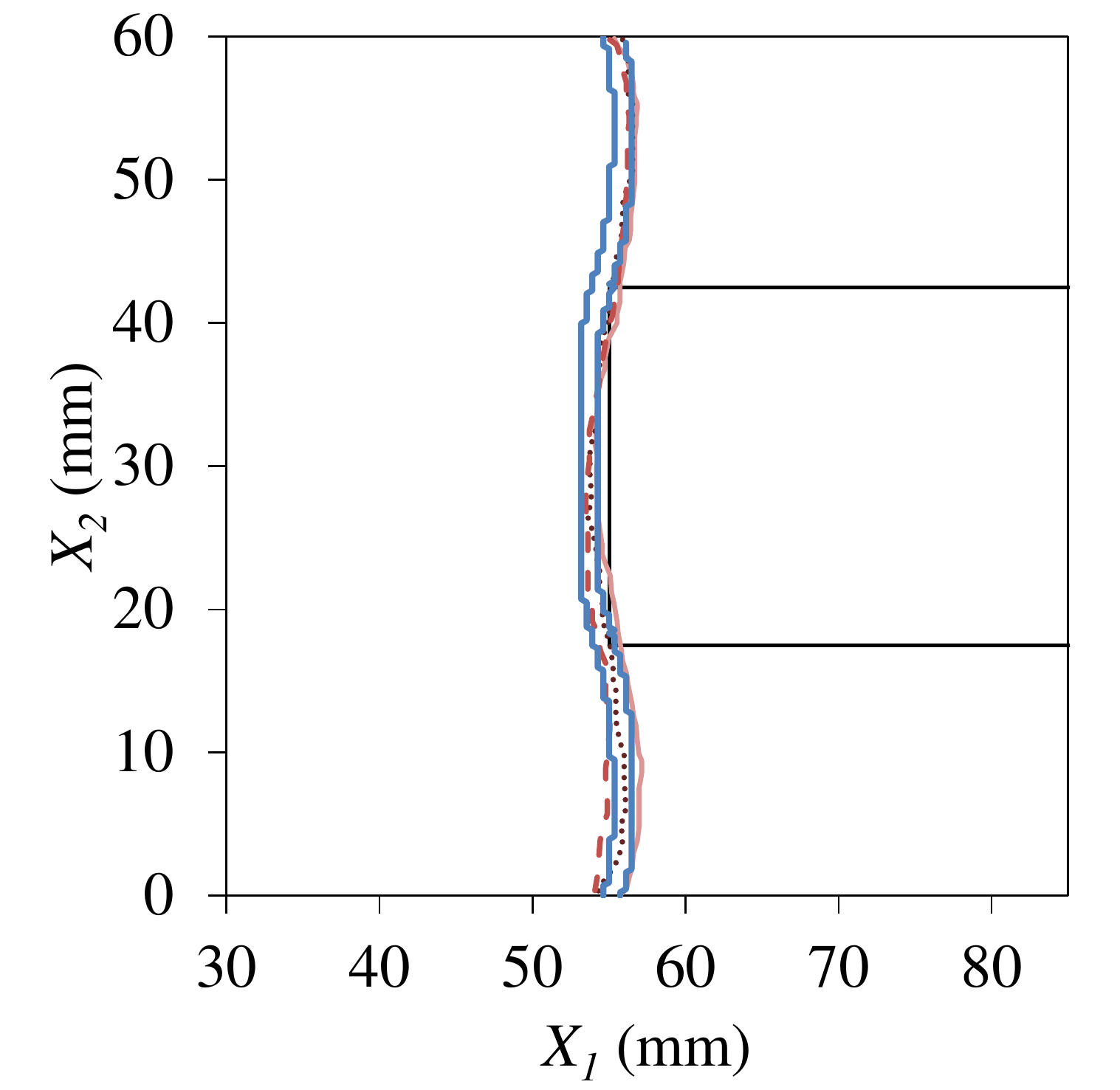} & \includegraphics[width=50mm]{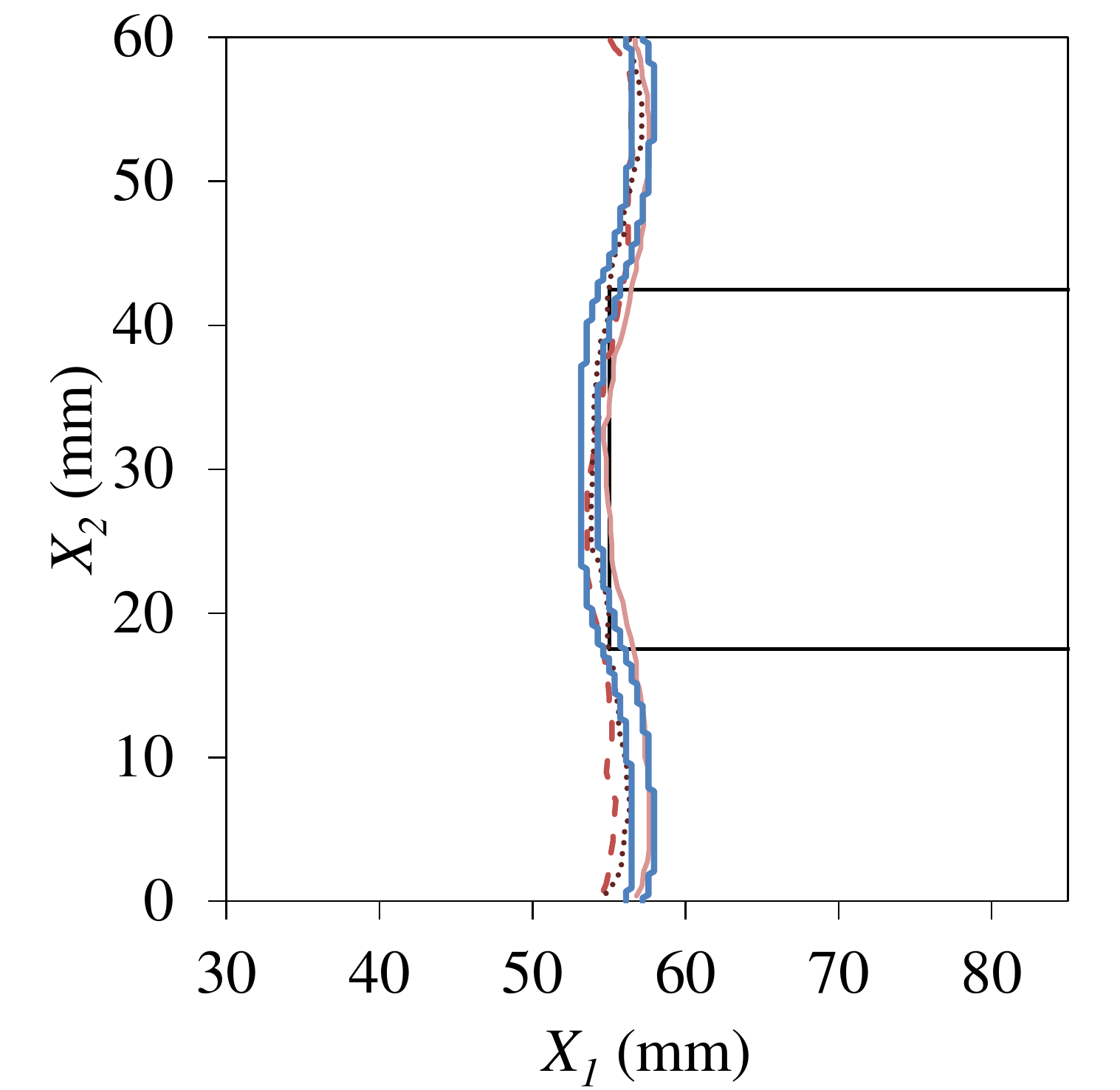} & \\
		\ \ \ \ \ \ g) Step 2. Fatigue loading & \ \ \ \ \ \ h) Step 2. Fatigue loading & \\
		\ \ \ \ \ \  $U_{3}^{max}= 5 $ mm 	& \ \ \ \ \ \ $U_{3}^{max}= 5$ mm  \\		
		\ \ \ \ \ \   $N= 80,000$ cycles	& \ \ \ \ \ \ $N= 410,000$ cycles & \\						  	
	\end{tabular}
	
	\caption{Comparison of numerical \cite{Carreras_data_num} and experimental \cite{demo_exp,Carreras_data} results for delamination front position: Steps 1 and 2.}
	\label{fig:Demo_fat}
\end{figure}

\begin{figure}[h]
	\centering
	\begin{tabular}{lll}
		\includegraphics[width=50mm]{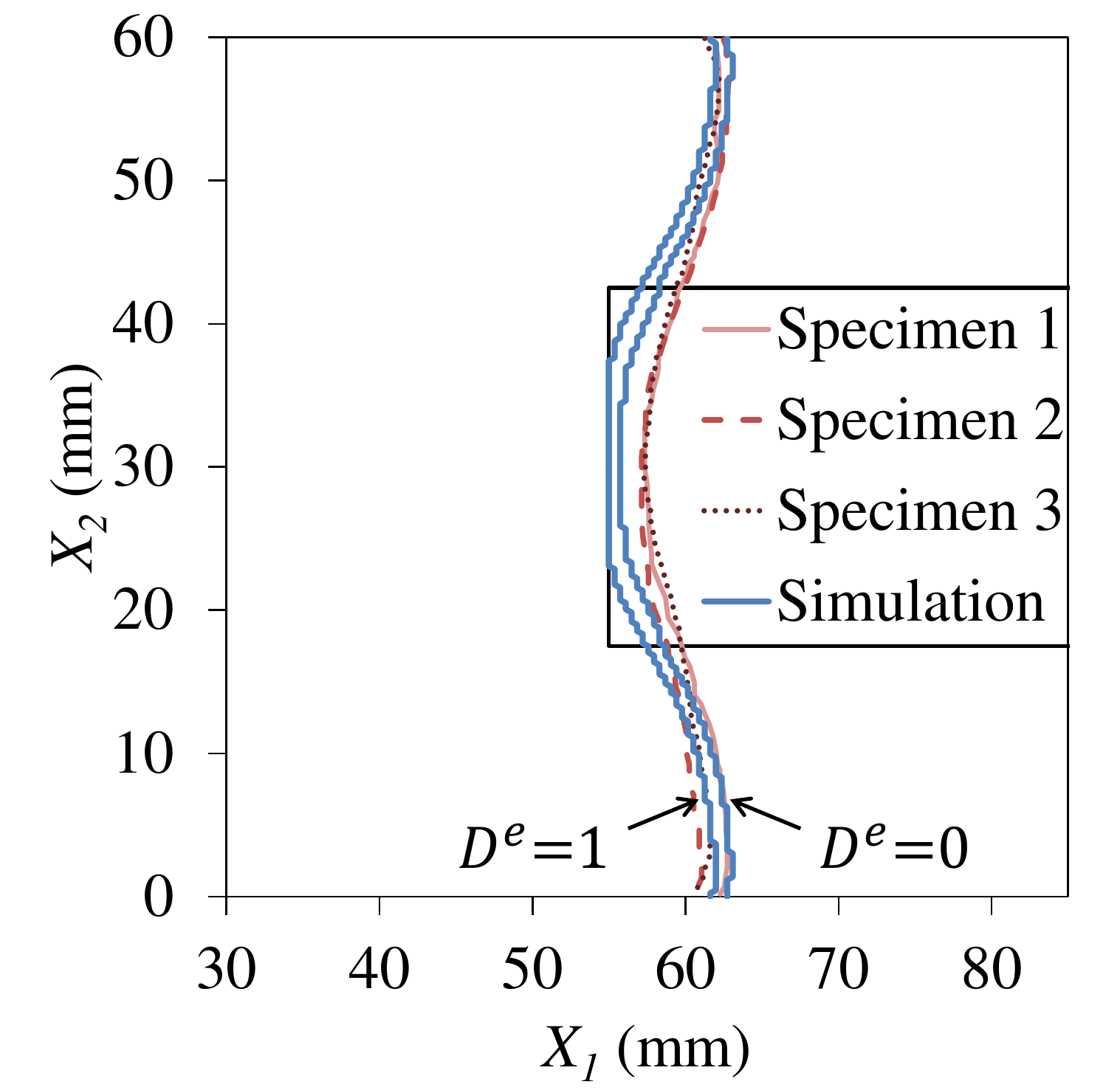} & \includegraphics[width=50mm]{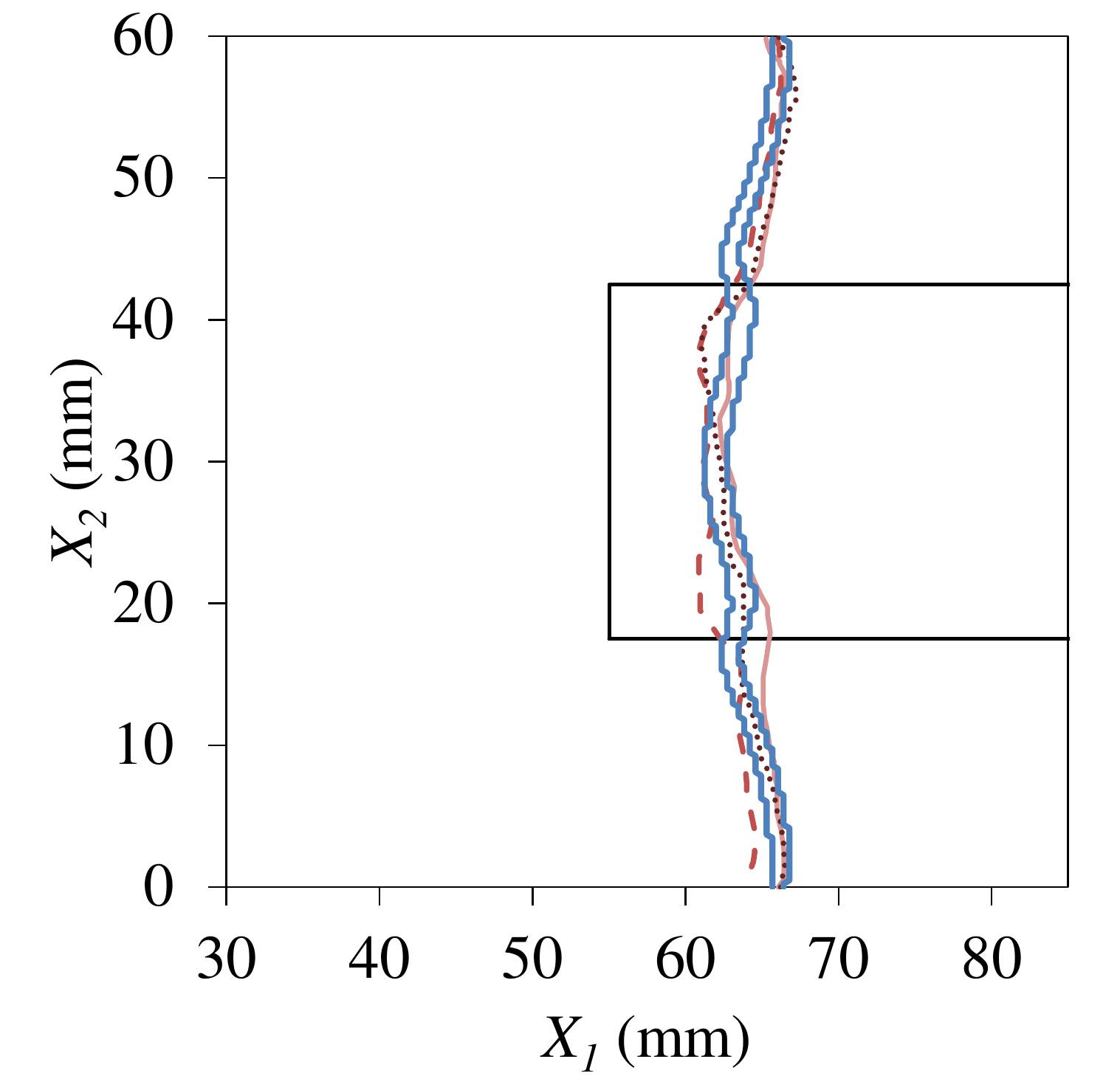}& \includegraphics[width=50mm]{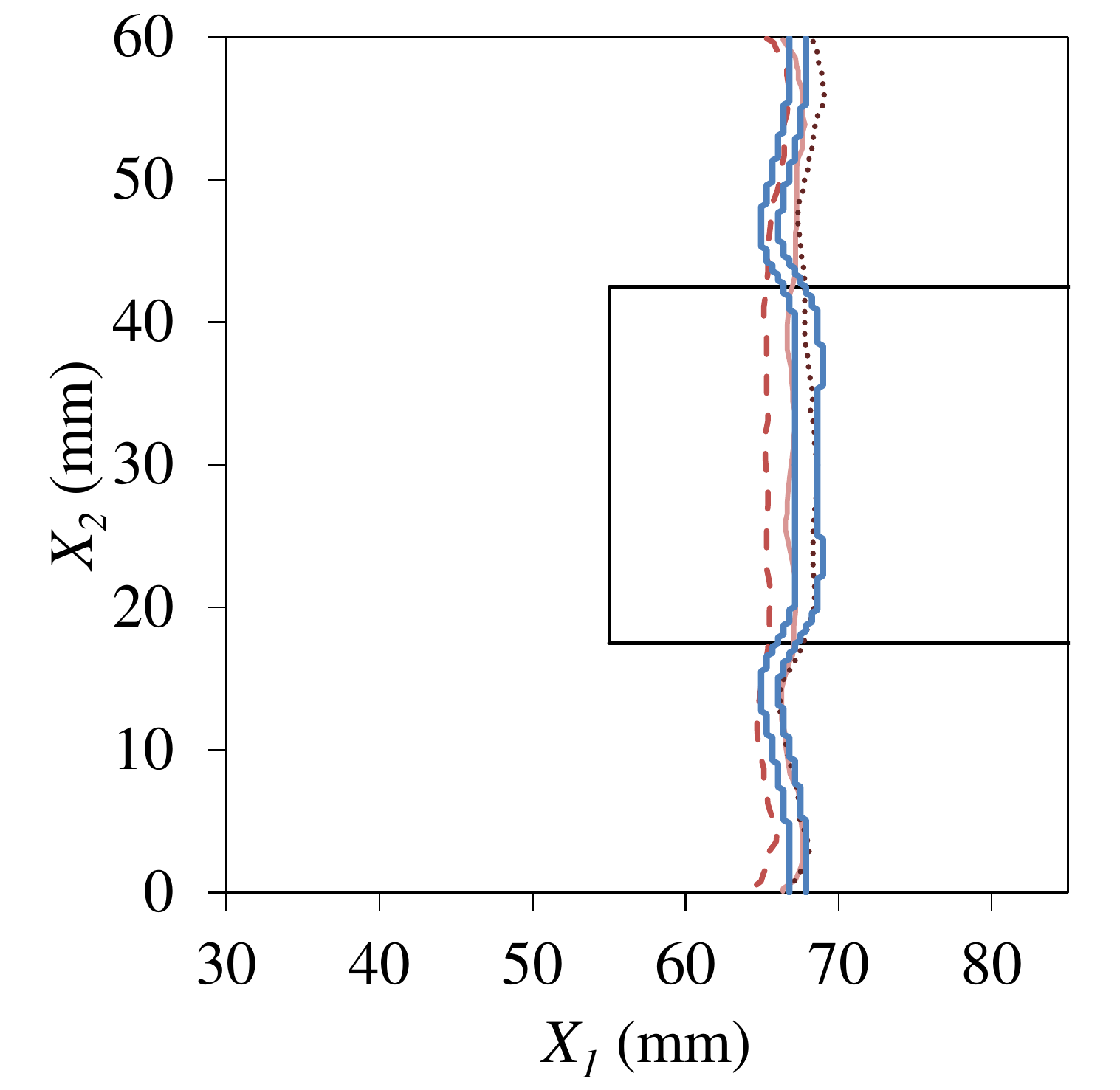} \\
		\ \ \ \ \ \ i) Step 3. Static loading & \ \ \ \ \ \ j) Step 4. Fatigue loading & \ \ \ \ \ \ k) Step 4. Fatigue loading \\
		\ \ \ \ \ \ $U_{3}= 10$ mm	& \ \ \ \ \ \ $U_{3}^{max}= 10 $ mm & \ \ \ \ \ \ $U_{3}^{max}= 10$ mm \\	
		& \ \ \ \ \ \ $N= 1,500$ cycles & \ \ \ \ \ \ $N= 3,000$ cycles\\	
		\includegraphics[width=50mm]{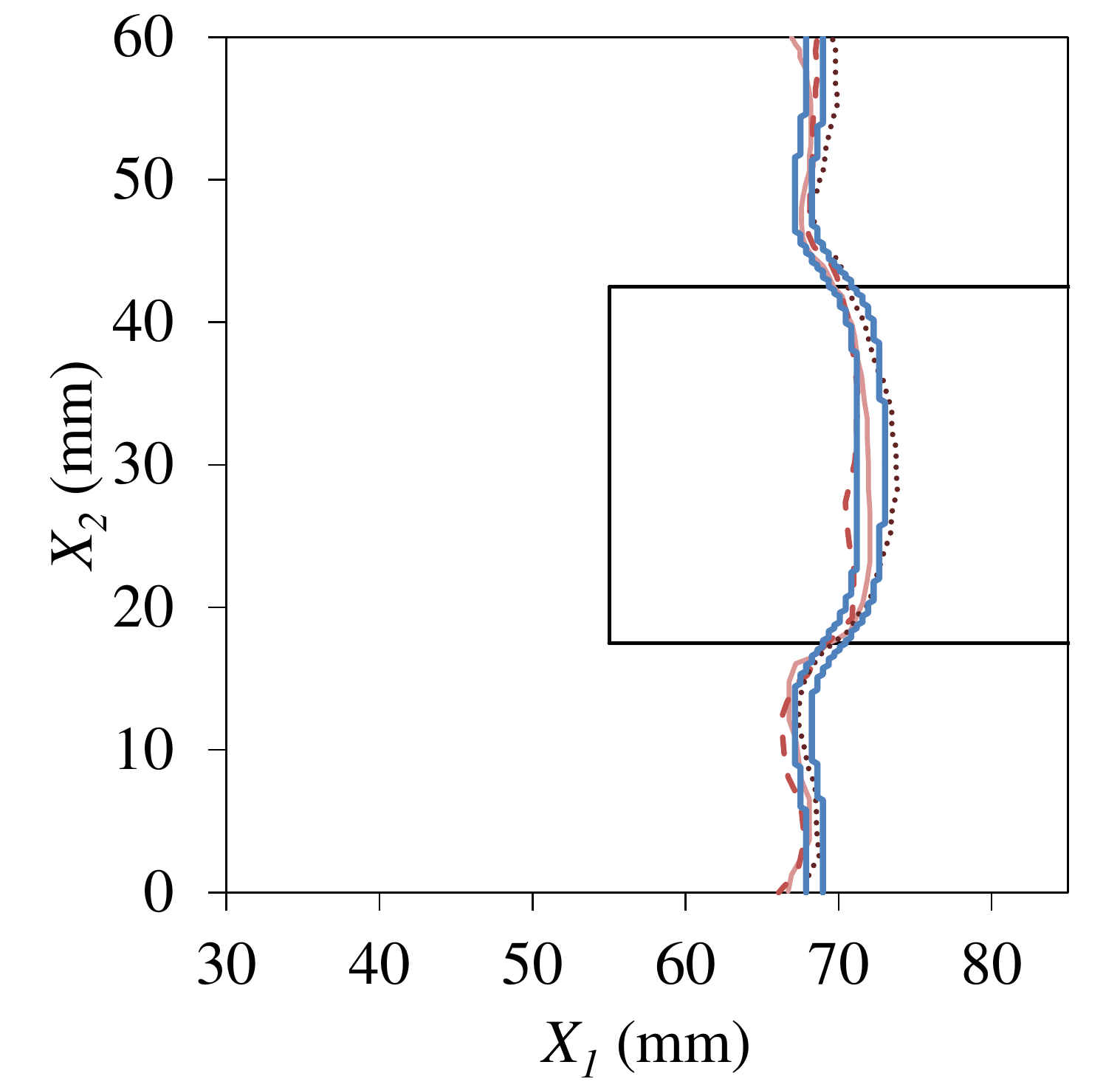} & \includegraphics[width=50mm]{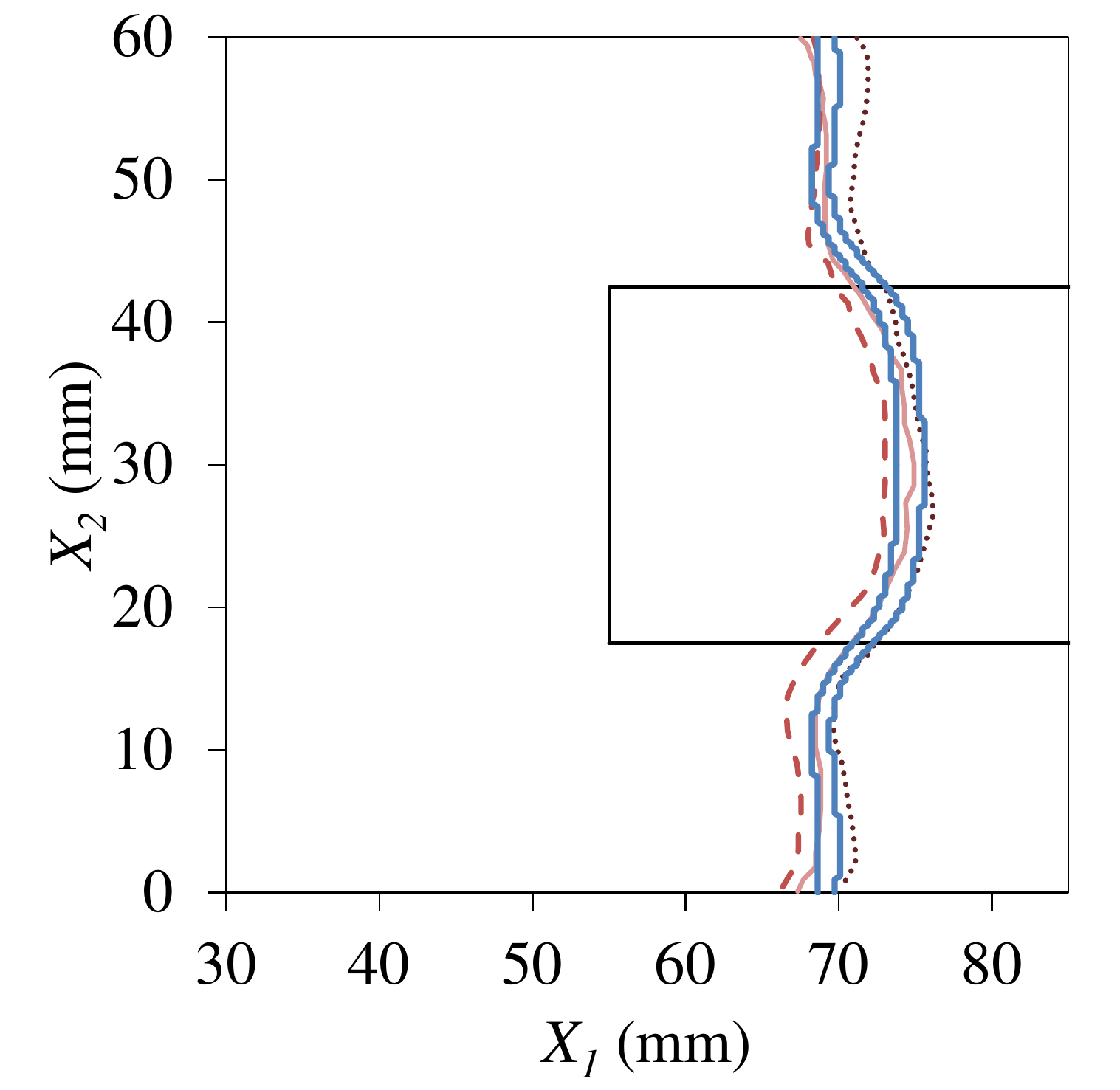} & \\
		\ \ \ \ \ \ l) Step 4. Fatigue loading & \ \ \ \ \ \ m) Step 4. Fatigue loading & \\
		\ \ \ \ \ \  $U_{3}^{max}= 10$ mm 	& \ \ \ \ \ \ $U_{3}^{max}= 10$ mm &  \\		
		\ \ \ \ \ \   $N= 5,000$ cycles	& \ \ \ \ \ \ $N= 10,000$ cycles &\\						  	
	\end{tabular}
	
	\caption{Comparison of numerical \cite{Carreras_data_num} and experimental \cite{demo_exp,Carreras_data} results for delamination front position: Steps 3 and 4.}
	\label{fig:Demo_fat2}
\end{figure}

In figures \ref{fig:Demo_fat} and \ref{fig:Demo_fat2}, the position of the delamination fronts of specimens 1, 2 and 3 is plotted for every stop done during the test. Due to the presence of micro-cracks, the fracture zone might contain some contrast liquid. Thus, it is hard to ensure whether the extracted delamination front location from the X-ray pictures corresponds to the beginning or the end of the fracture process zone. For this reason, both the 1 and 0-damage isolines delimiting the cohesive zone are plotted in figures \ref{fig:Demo_fat} and \ref{fig:Demo_fat2} for the comparison with results from experiments.

It can be observed that the delamination front shape and growth rate with the number of cycles is reproduced with high accuracy by the proposed simulation tool. In figures \ref{fig:Demo_fat}.a-\ref{fig:Demo_fat}.h, the delamination front shape changes from initial straight to curved front shape as the delamination approaches the stiffened region. Under fatigue loading, the front is arrested at the reinforcement edges due to the supplied stiffness. It is observed in Figure \ref{fig:Demo_fat}.h that the delamination front of specimens 1, 2 or 3 did not surpass the reinforcement edges within 410,000 cycles. Delamination arrestment is also captured by the simulation. Then, during the quasi-static increment of displacement in Step 3, the delamination front surpasses the reinforcement edge and keeps advancing under the reinforcement during fatigue loading in Step 4. During fatigue Step 4, the crack front shape changes once again.

\begin{figure}[h!]
	\centering
	\begin{adjustbox}{width=0.85\textwidth}
		\begin{tabular}{cc}
			
			\includegraphics[width=68mm]{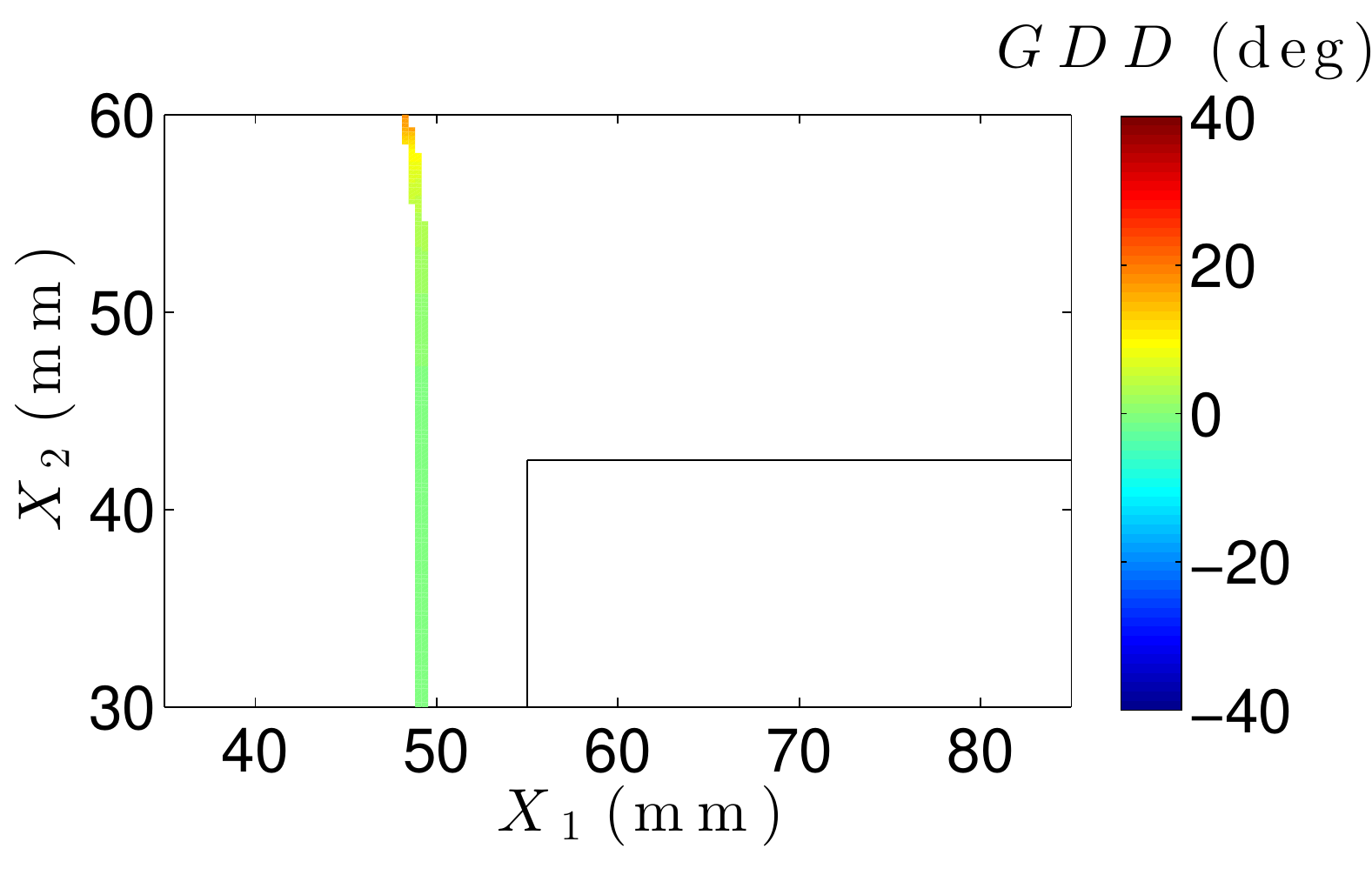} & \includegraphics[width=68mm]{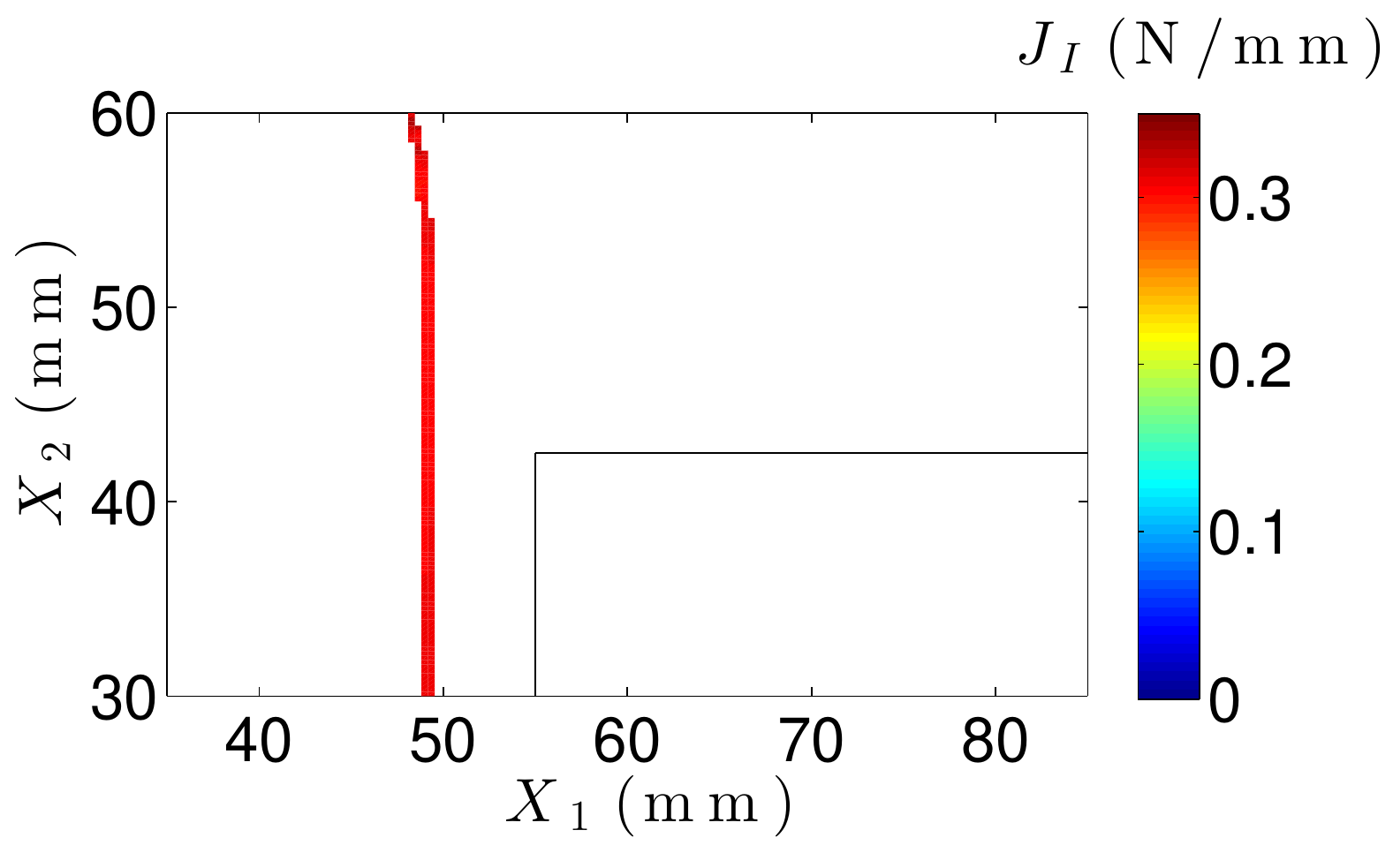} \\
			\multicolumn{2}{c}{a) Step 1. Static loading $U_{3}= 5$ mm}\\
			\includegraphics[width=68mm]{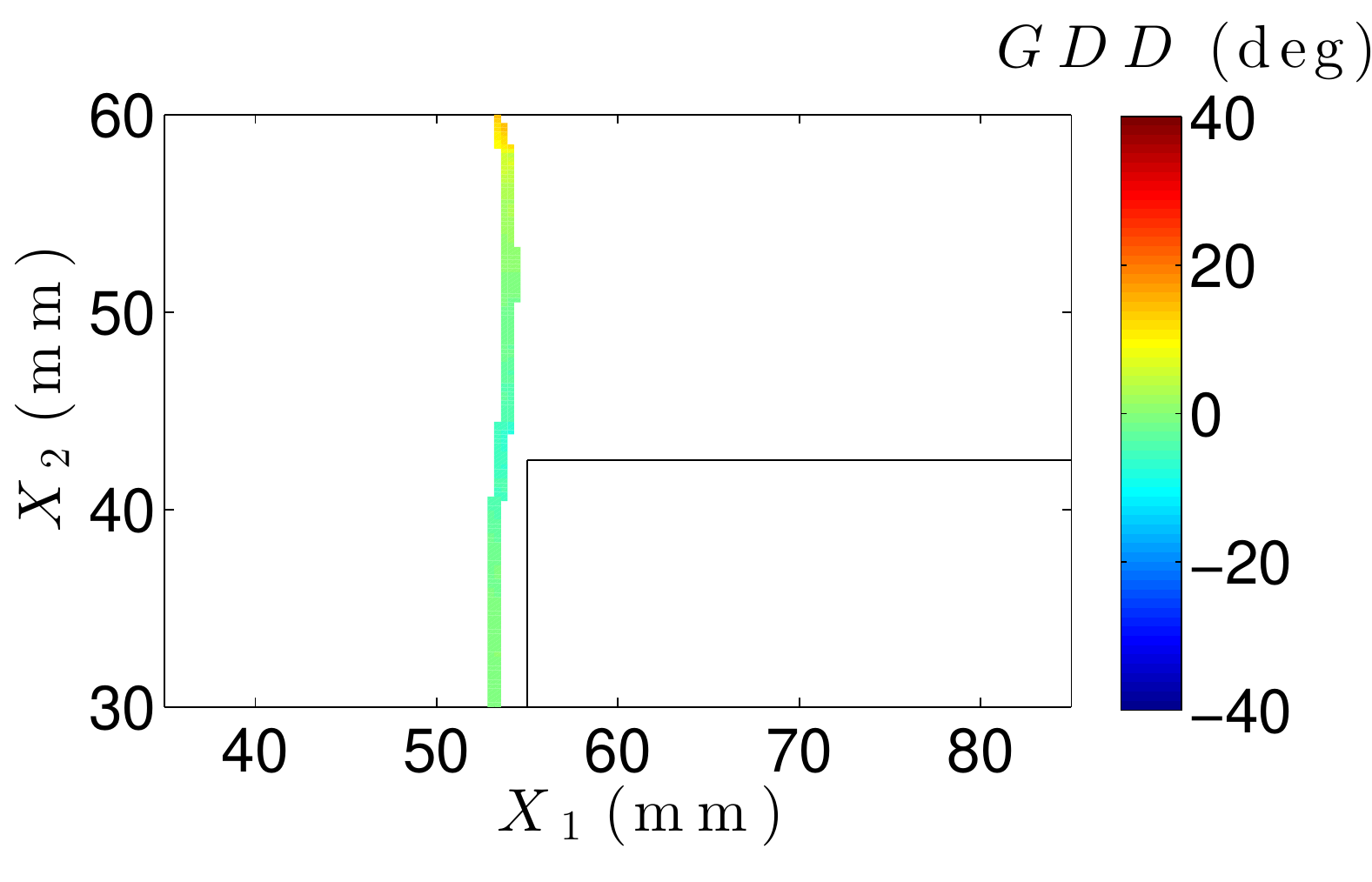} & \includegraphics[width=68mm]{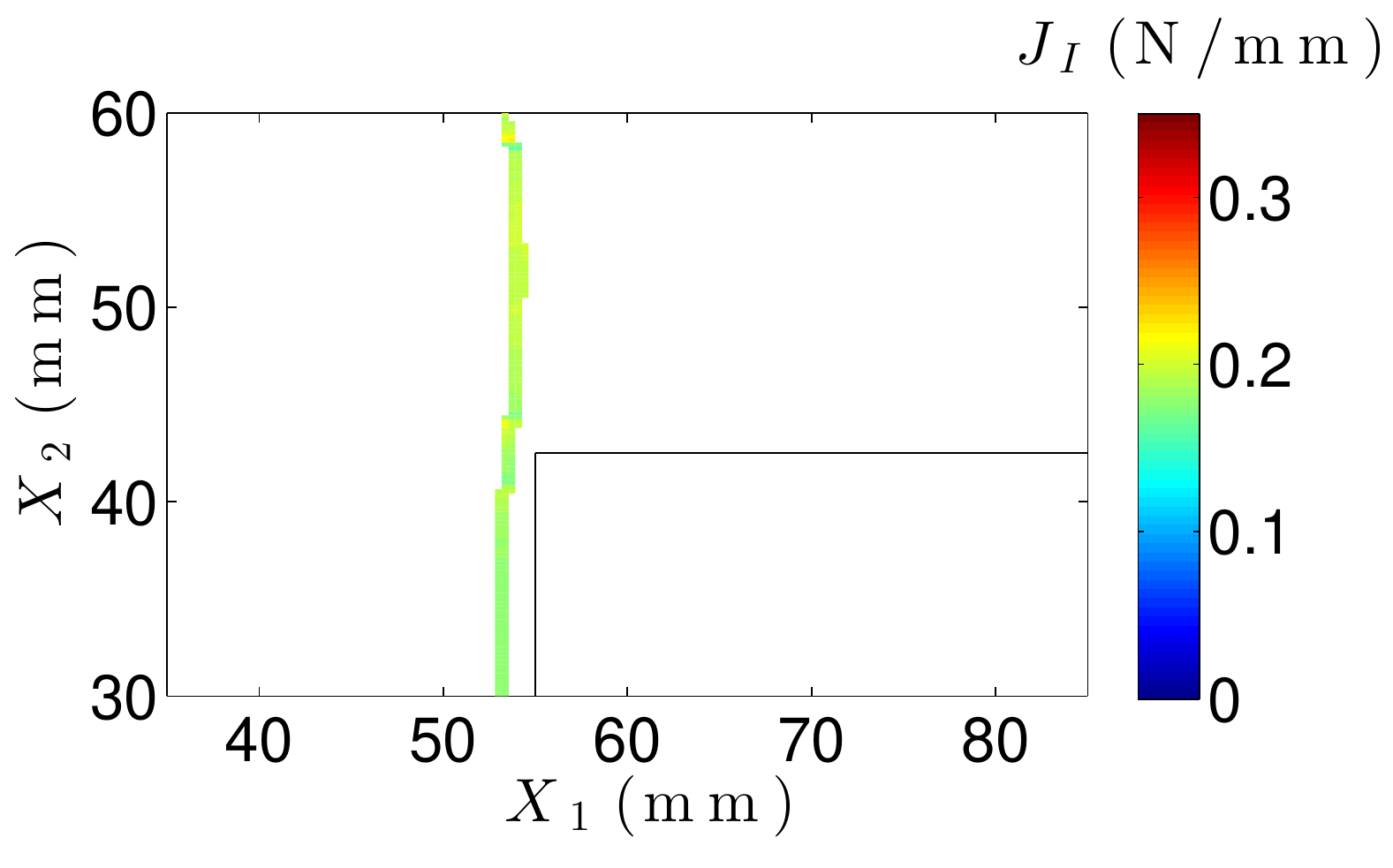} \\
			\multicolumn{2}{c}{b) Step 2. Fatigue loading $U_{3}^{max}= 5$ mm $N= 3,000$ cycles}\\				
			\includegraphics[width=68mm]{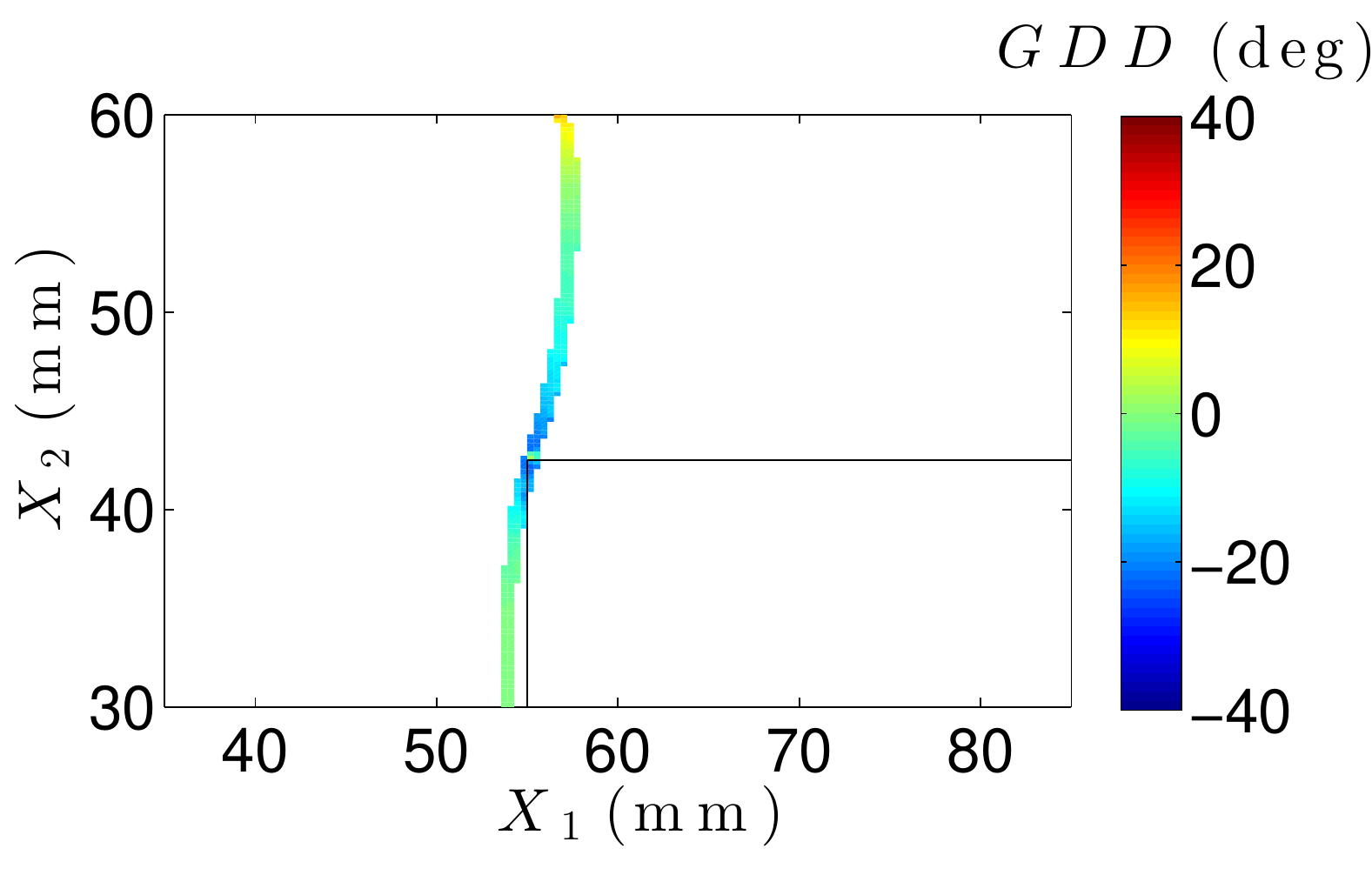} & \includegraphics[width=68mm]{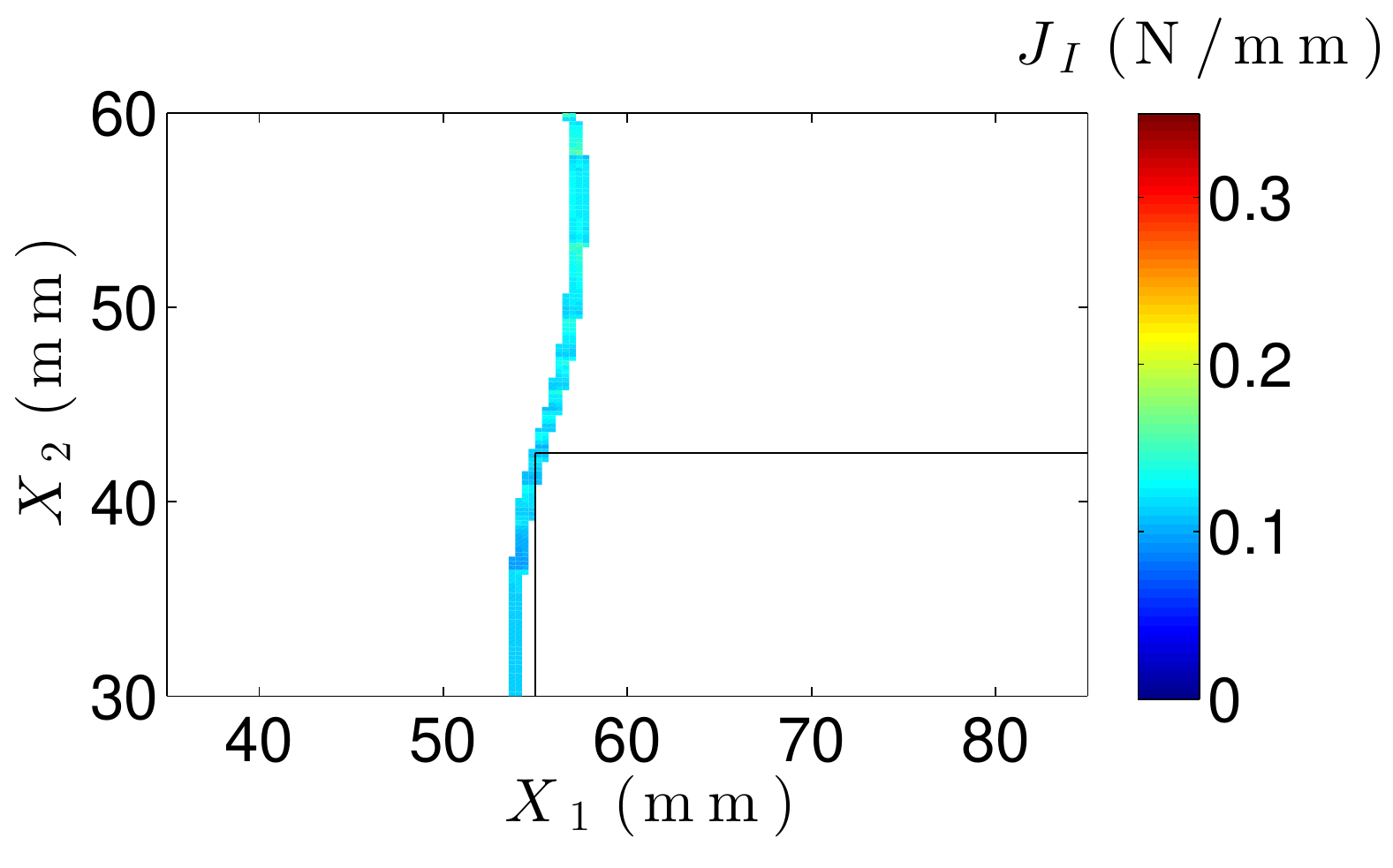} \\	
			\multicolumn{2}{c}{c) Step 2. Fatigue loading $U_{3}^{max}= 5$ mm $N= 410,000$ cycles}\\				
		\end{tabular}
	\end{adjustbox}	
	\caption{Historical evolution of growth driving direction (GDD) and mode I component of the $J$-integral: Steps 1 and 2. The mode II and III components of the $J$-integral are negligible.}
	\label{fig:Sim_resul}
\end{figure}

\begin{figure}[h!]
	\centering
	\begin{adjustbox}{width=0.85\textwidth}
		\begin{tabular}{cc}
			\includegraphics[width=68mm]{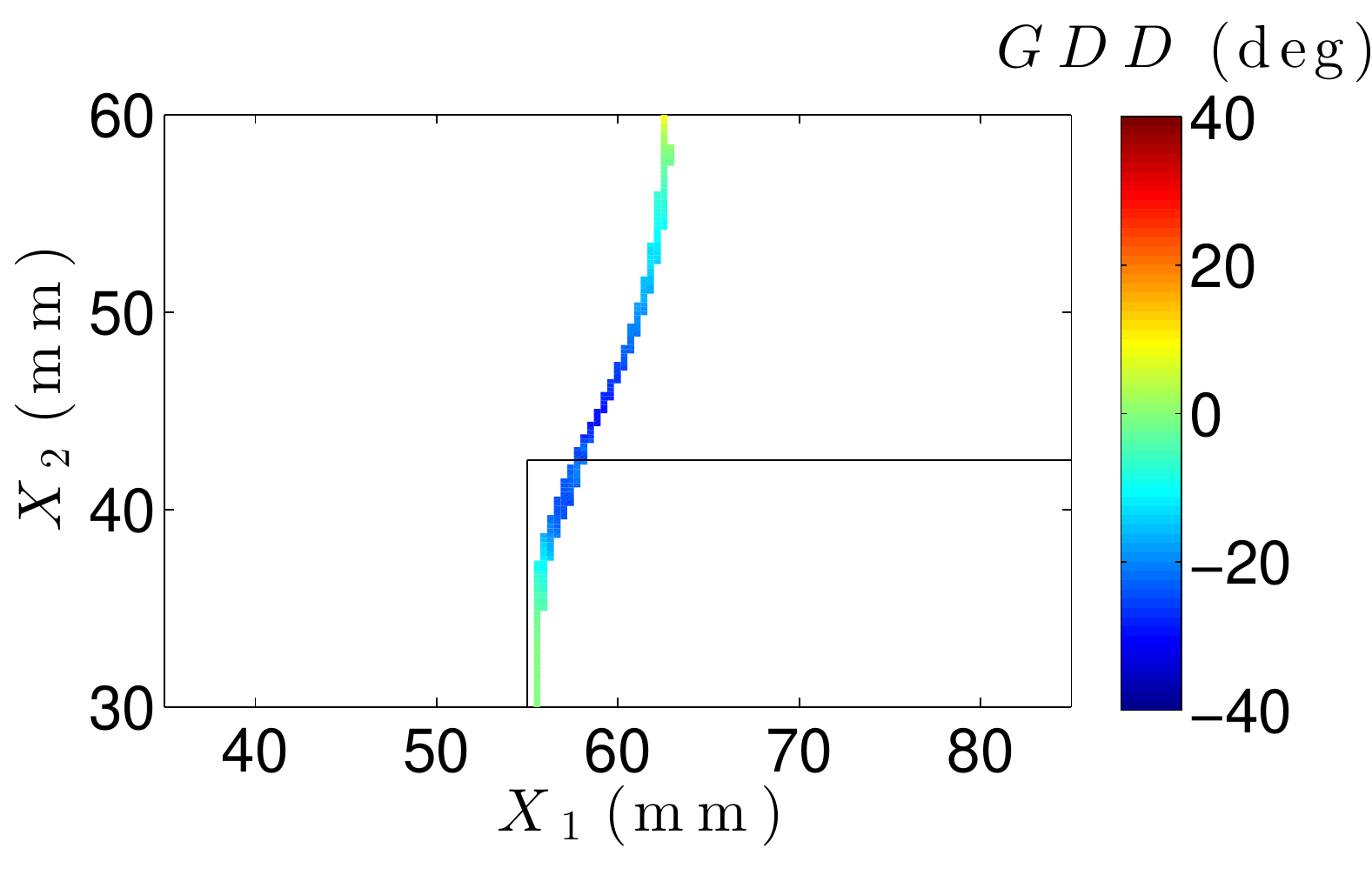} & \includegraphics[width=68mm]{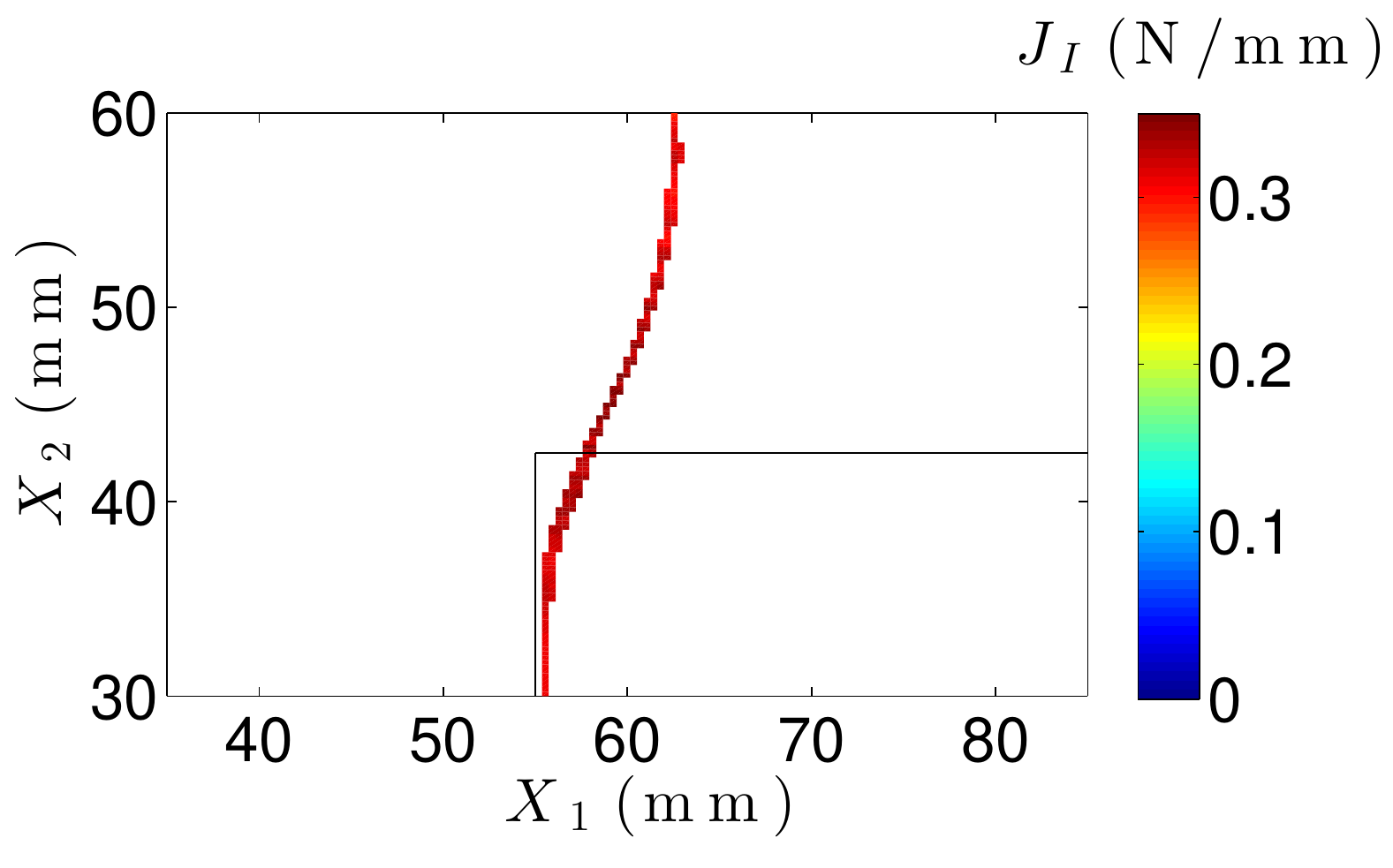} \\
			\multicolumn{2}{c}{d) Step 3. Static loading $U_{3}= 10$ mm}\\		
			\includegraphics[width=68mm]{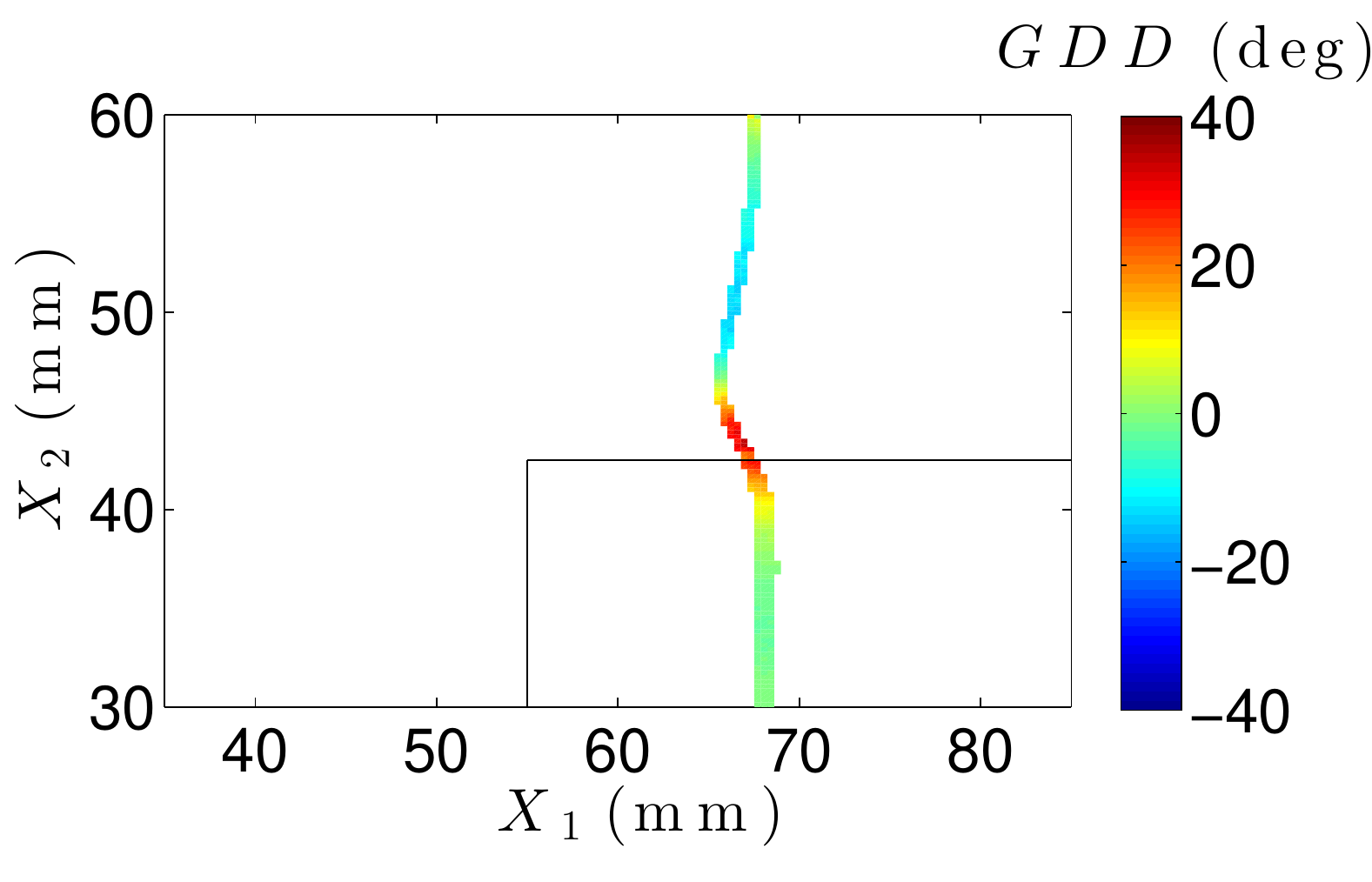} & \includegraphics[width=68mm]{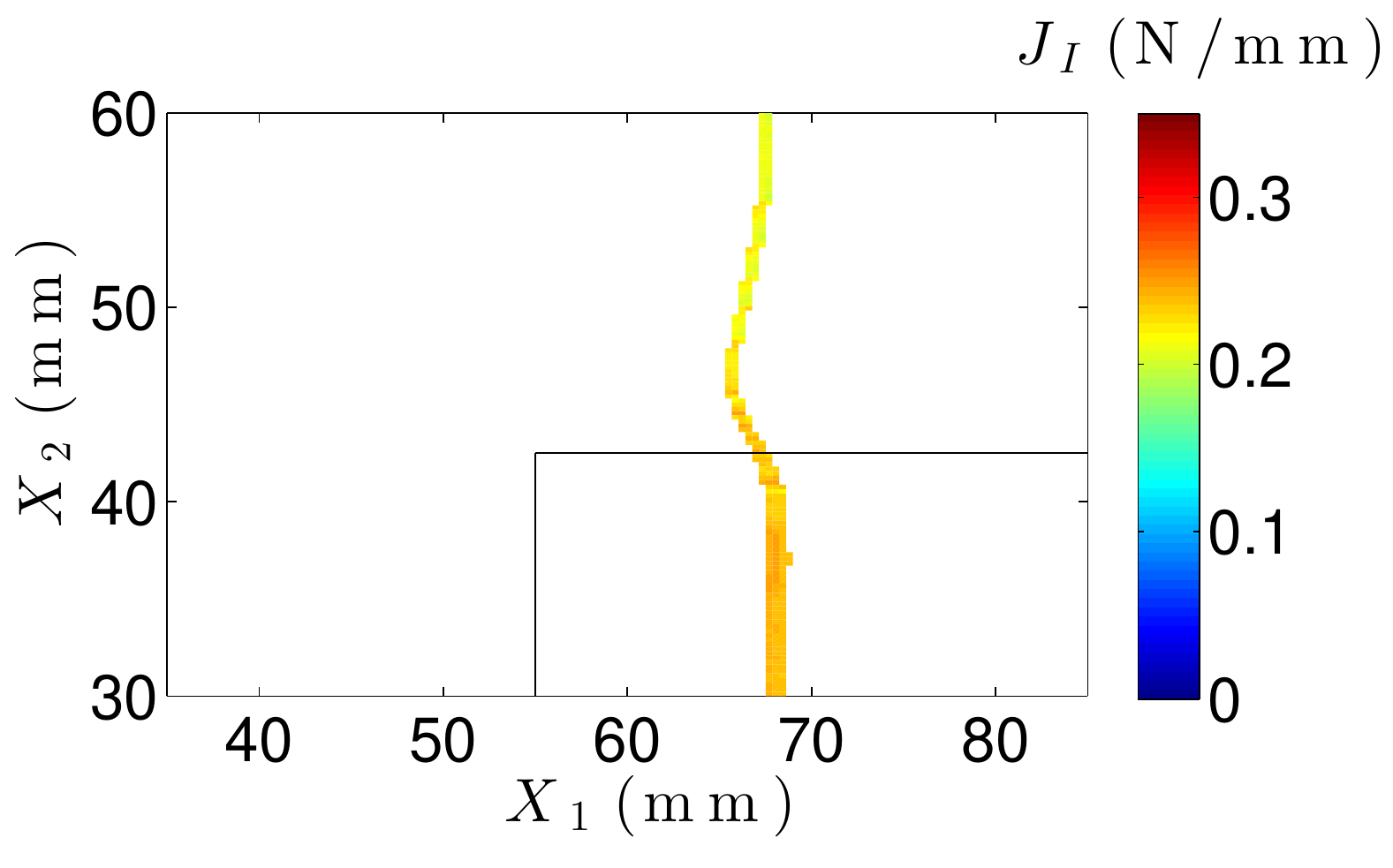} \\
			\multicolumn{2}{c}{e) Step 4. Fatigue loading $U_{3}^{max}= 10$ mm $N= 3,000$ cycles}\\		
			\includegraphics[width=68mm]{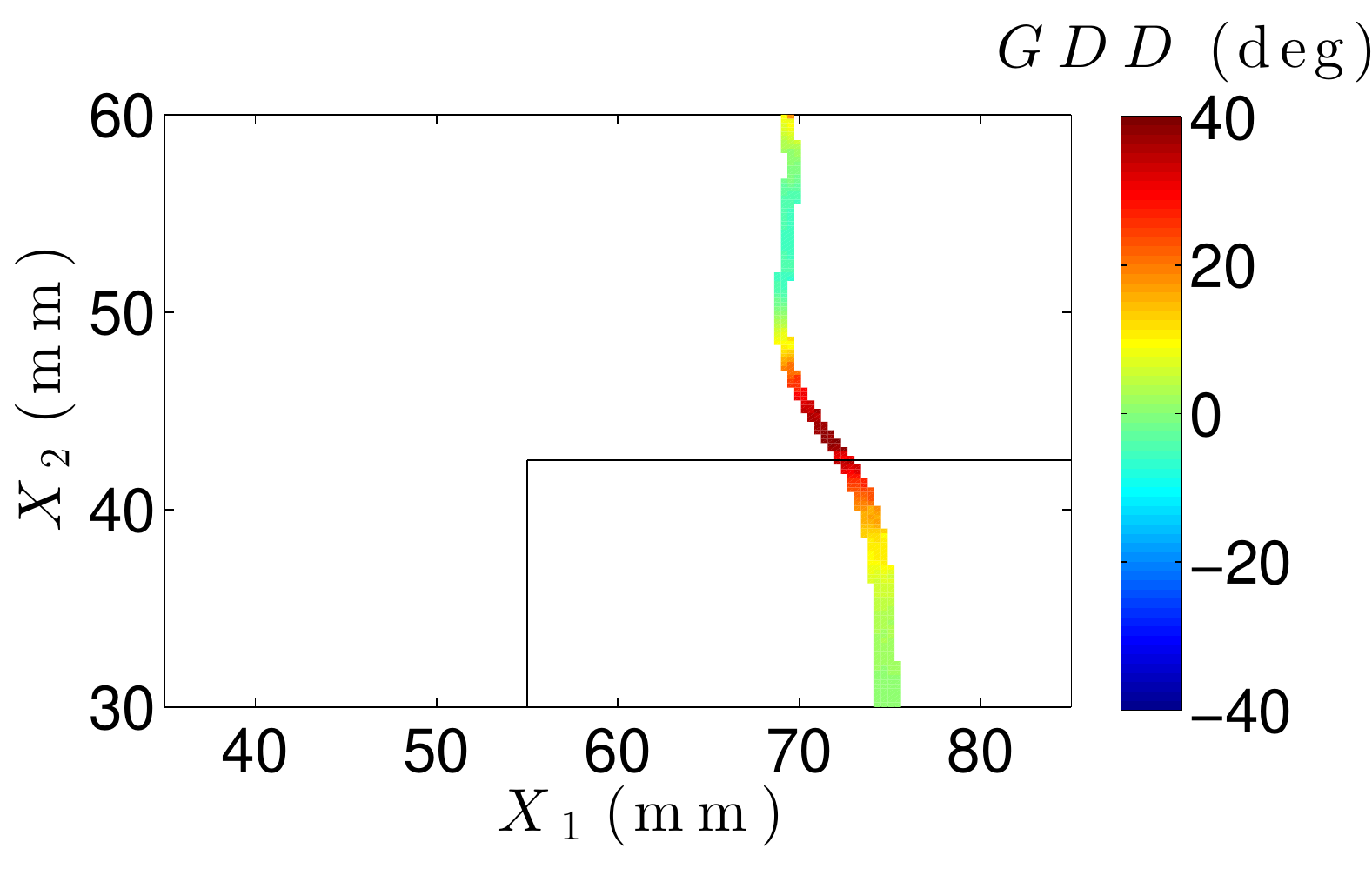} & \includegraphics[width=68mm]{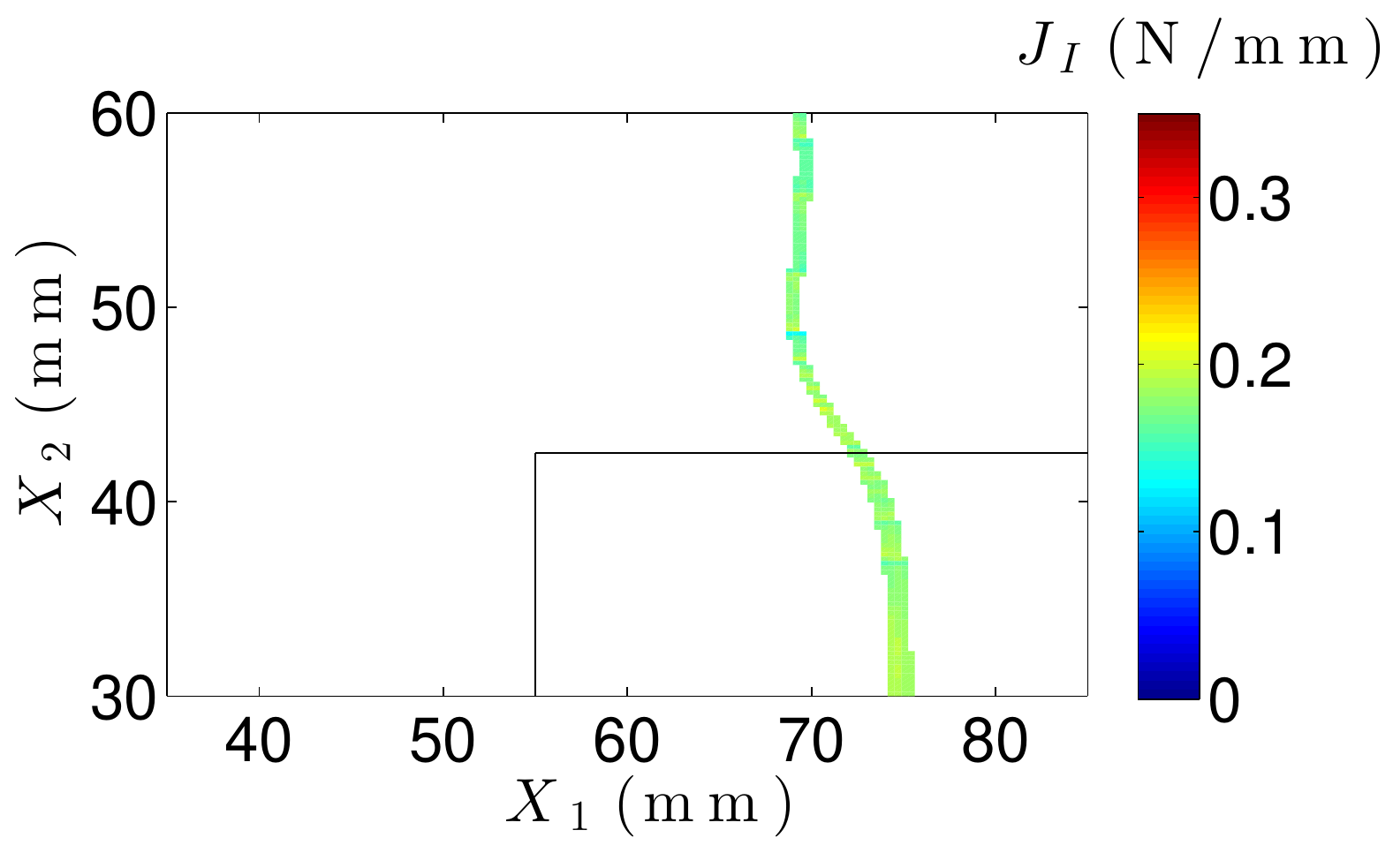} \\
			\multicolumn{2}{c}{f) Step 4. Fatigue loading $U_{3}^{max}= 10$ mm $= 10,000$ cycles}\\			
		\end{tabular}
	\end{adjustbox}		
	\caption{Historical evolution of growth driving direction (GDD) and mode I component of the $J$-integral: Steps 3 and 4. The mode II and III components of the $J$-integral are negligible.}
	\label{fig:Sim_resul2}
\end{figure}

The historical evolution of the growth driving direction (GDD) and the mode I component of the $J$-integral are plotted in figures \ref{fig:Sim_resul} and \ref{fig:Sim_resul2}. The mode II and III components of the $J$-integral are not plotted since they are negligible under these loading conditions. The results are evaluated within the cohesive zone and projected on the mid-surface. It can be visually inspected that the values for the GDD correspond to the angle of the normal to the delamination front with the global $X_{1}$-axis. More accurate demonstrations of the capabilities of the GDD formulation are provided in \cite{Carreras2018}. In Figure  \ref{fig:Sim_resul}.a, it can be observed that the computed $J$-integral during quasi-static propagation (Step 1) is equal to the mode I fracture toughness, $\mathcal{G}_{Ic}$. Then, during Step 2 (fatigue loading with constant maximum cyclic displacement), the crack front is advancing with a $J$-value lower than the fracture toughness, $\mathcal{G}_{Ic}$, and $J$ is reducing as the crack front advances (c.f. figures \ref{fig:Sim_resul}.b and \ref{fig:Sim_resul}.c). The same is repeated for steps 3 and 4: in Figure \ref{fig:Sim_resul2}.d, the $J$-integral during quasi-static propagation again equals the mode I fracture toughness, $\mathcal{G}_{Ic}$, whereas, in figures \ref{fig:Sim_resul2}.e and \ref{fig:Sim_resul2}.f, $J$ decreases with crack propagation. The accuracy of the 3D CZ $J$-integral in the computation of the mode-decomposed energy release rate is further demonstrated in \cite{Carreras_Jint}.

\section {Summary and conclusions} \label{sec:Conclusions}

In this work, a new simulation method for high-cycle fatigue-driven delamination has been proposed. The method is applicable to 3D analysis of delamination in layered composite structures. A cohesive zone model (CZM) approach is used. The model is capable to account for a non-negligible fracture process zone and arbitrarily-shaped delamination fronts avoiding the need of using re-meshing techniques. Moreover, the model is based on a Paris' law-based expression, as most of the current phenomenological models for representing crack growth rate observations. 

Recent developments by the authors, like the formulation for the efficient identification of the growth driving direction (GDD) and the computation of the mode-decomposed $J$-integral in 3D analysis (3D CZ $J$-integral), have been integrated to this numerical tool. First, the GDD formulation is used to compute the spatial derivatives of some quantities of the cohesive zone in the direction of crack advance. This enhances the accuracy and robustness of the method as it avoids the introduction of any fitting parameter. Secondly, the 3D CZ $J$-integral is used to evaluate the mode-decomposed energy release rate that feeds the expression for the crack growth rate. By applying this formulation, a macroscopic measure of the energy release rate is obtained, which takes into account the current traction-displacement jump field of the entire cohesive zone. To the authors knowledge, this has not been previously addressed in 3D analysis using a CZM approach.

The method has been implemented in an FE framework as a user-defined cohesive element in Abaqus. The capability to reproduce the Paris' law curve under different loading conditions of mode mixity and load ratio has been demonstrated. Moreover, in order to assess its predictive capabilities in specimens that cannot be simplified to 2D representations, it has been evaluated against a benchmark case with varying crack growth rate and front shape. In all cases, the documented discrepancy between experimental and simulation data is very small.

The method proposed is a key step in providing a simulation tool that contributes to reduce the amount of experimental testing needed in the dimensioning of real layered components and structures undergoing fatigue delamination. Further work on this method is ongoing and embraces enhancing the model capabilities to include fiber bridging effects and R-curve behavior on d$a$/d$N$ \cite{yao2014bridging,yao2016effect}. The formulation presented is compatible for its use with other cohesive zone law shapes that are better suited to model delamination with large scale bridging and that can be obtained from experiments \cite{Simon_multi,Simon_inverse}. In addition, due to the fact that the method is based on the concept of the GDD, the formulation may, in future works, be enhanced to include in-plane anisotropy with respect to fracture properties, i.e. $\mathcal{G}_{c}$ and d$a$/d$N$, which may depend on the off-axis angle between the lamina orientation and the crack growth direction \cite{Lindgaard_angle}.

\section{Acknowledgements}

This work has been partially funded by the European Union by the financial supports of ERANet AirTN 01/2013 under the project entitled ``Methodology to design composite structures resistant to intra- and interlaminar damage (static \& fatigue)- MERINDA'' and by the Spanish Government (Ministerio de Economia y Competitividad) under contract TRA2015-71491-R, cofinanced by the European Social Fund. In addition, this project has received funding from the European Union's Horizon 2020 research and innovation programme under grant agreement No 763990.

\section{Data availability}
The processed data required to reproduce these findings are available to download from

\noindent http://dx.doi.org/10.17632/v7bgzx9gmw.1 and http://dx.doi.org/10.17632/87r49xbrp3.3, open-source online data repositories hosted at Mendeley Data \cite{Carreras_data,Carreras_data_num}.



\bibliographystyle{elsarticle-num}
\bibliography{library}

\end{document}